\documentclass[twocolumn,showpacs,prb,aps,amsmath,amssymb,superscriptaddress]{revtex4}

\usepackage{graphicx}% Include figure files
\usepackage{bm}% bold math
\usepackage{color}

\begin{document}

\title{
Magnetic phase diagram of the spin-$1/2$ antiferromagnetic zigzag ladder
}
\author{Toshiya Hikihara}
\affiliation{Department of Physics, Hokkaido University,
Sapporo 060-0810, Japan}
\author{Tsutomu Momoi}
\author{Akira Furusaki}
\affiliation{Condensed Matter Theory Laboratory, RIKEN,
Wako, Saitama 351-0198, Japan}
\author{Hikaru Kawamura}
\affiliation{Department of Earth and Space Science,
Faculty of Science, Osaka University, Toyonaka 560-0043, Japan}
\date{\today}

\begin{abstract}
We study the one-dimensional spin-1/2 Heisenberg model with
antiferromagnetic nearest-neighbor $J_1$ and next-nearest-neighbor $J_2$
exchange couplings in magnetic field $h$.
With varying dimensionless parameters $J_2/J_1$ and $h/J_1$,
the ground state of the model exhibits several phases
including three gapped phases (dimer, $1/3$-magnetization plateau, and
fully polarized phases)
and four types of gapless Tomonaga-Luttinger liquid (TLL) phases
which we dub TLL1, TLL2, spin-density-wave (SDW$_2$),
and vector chiral phases.
From extensive numerical calculations
using the density-matrix renormalization-group method,
we investigate various (multiple-)spin correlation functions
in detail, and determine dominant and subleading correlations in each phase.
For the one-component TLLs, i.e., the TLL1, SDW$_2$, and vector chiral phases,
we fit the numerically obtained correlation functions
to those calculated from effective low-energy theories of TLLs,
and find good agreement between them.
The low-energy theory for each critical TLL phase is thus identified,
together with TLL parameters which control
the exponents of power-law decaying correlation functions.
For the TLL2 phase, we develop an effective low-energy theory
of two-component TLL
consisting of two free bosons (central charge $c=1+1$),
which explains numerical results of
entanglement entropy and Friedel oscillations of local magnetization.
Implications of our results to possible magnetic phase transitions
in real quasi-one-dimensional compounds are also discussed.
\end{abstract}


\maketitle

\section{Introduction}\label{sec:Intro}

Frustrated quantum antiferromagnets have
long been a subject of active research,
since Anderson\cite{Anderson} suggested resonating-valence-bond
ground state for a triangular lattice antiferromagnet.
Recent experimental studies of quasi-two-dimensional compounds,
such as the organic Mott insulator\cite{Shimizu}
$\kappa$-(BEDT-TTF)$_2$Cu$_2$(CN)$_3$ and
the transition metal chloride Cs$_2$CuCl$_4$,\cite{Coldea}
have further prompted theoretical research of
anisotropic triangular lattice antiferromagnets.\cite{Watanabe04,Motrunich05,Lee2,Alicea06,Veillette05,StarykhB2007,Kawakami09,Kohno}
In these anisotropic quasi-two-dimensional antiferromagnets
combination of frustrated exchange interactions
and strong quantum fluctuations
suppresses tendency toward conventional magnetic orders,
thereby opening up possibilities of exotic quantum states.

A zigzag spin ladder is a one-dimensional (1D) strip of the anisotropic
triangular lattice spin system, and can be regarded as a minimal, toy model
of (strongly anisotropic quasi-two-dimensional) frustrated quantum magnets.
Furthermore,
the 1D $J_1$-$J_2$ Heisenberg model on the zigzag ladder is in itself
a good model for various quasi-1D magnetic compounds,
such as (N$_2$H$_5$)CuCl$_3$,\cite{Brown1979,Hagiwara2001,Maeshima2003}
Rb$_2$Cu$_2$Mo$_3$O$_{12}$,\cite{Hase2004}
and LiCuVO$_4$.\cite{Enderle2005,Banks2007,Buttgen2007,Naito2007,Schrettle2008}
Despite its simplicity, the 1D $J_1$-$J_2$ Heisenberg model has been
shown to
exhibit various unconventional phases
under magnetic field (as we summarize
below).\cite{OkunishiT2003,HikiharaKMF2008,SudanLL2008,Heidrich-Meisner2009}
In this paper we
aim to clarify the nature of the phases
in the ground-state phase diagram of the 1D
spin-$1/2$ $J_1$-$J_2$ Heisenberg model
under magnetic field,
when
both the nearest- and next-nearest-neighbor exchange couplings are
\textit{antiferromagnetic} (AF).
To this end, we study in detail spin correlations in each phase
using the numerical density matrix renormalization group (DMRG) method
as well as low-energy effective theory based on bosonization.

The Hamiltonian of the $J_1$-$J_2$ Heisenberg zigzag spin ladder is given by
\begin{equation}
\mathcal{H} = J_1 \sum_l {\bm s}_l \cdot {\bm s}_{l+1}
+ J_2 \sum_l {\bm s}_l \cdot {\bm s}_{l+2}
-h \sum_l s^z_l,
\label{eq:Ham}
\end{equation}
where ${\bm s}_l$ is a spin-$1/2$ operator at $l$th site,
$J_1$ and $J_2$ are respectively the nearest- and next-nearest-neighbor
exchange couplings ($J_1>0$ and $J_2>0$),
and $h$ is external magnetic field along the $z$-direction.

In the classical limit, the ground state of
the zigzag ladder $J_1$-$J_2$
Heisenberg antiferromagnet has a helical magnetic structure
\begin{equation}
{\bm s}_l =
s (\sin \theta^c \cos \phi_l^c,\sin \theta^c \sin \phi_l^c,\cos \theta^c)
\end{equation}
with a pitch angle
\begin{equation}
\phi^c = \phi_{l +1}^c -\phi_l^c
 = \pm \arccos\left(\frac{-J_1}{4J_2}\right)
\end{equation}
and a canting angle
\begin{equation}
\theta^c =
 \arccos\left(\frac{4hJ_2}{s(J_1 + 4J_2)^2}\right)
\end{equation}
for $J_2/J_1>1/4$, whereas the ground state has canted
N\'eel order for $J_2/J_1 \le 1/4$.

In the quantum ($s=1/2$) case,
the ground-state properties of the model (\ref{eq:Ham}) change drastically
from the classical spin state.
The ground state at zero magnetic field $h=0$ has been understood quite well.
For small $J_2/J_1 < (J_2/J_1)_{\rm c}$,
the ground state is in
a critical Tomonaga-Luttinger liquid (TLL) phase
with gapless excitations.
The model undergoes a quantum phase transition
at $(J_2/J_1)_{\rm c} = 0.2411$,\cite{JullienH1983,OkamotoN1992,Eggert1996}
to a gapped phase with spontaneous
dimerization\cite{MajumdarG1969A,MajumdarG1969B,Haldane1982,WhiteA1996}
for $J_2/J_1 > (J_2/J_1)_{\rm c}$.
It is also known that the model exhibits a long-range order (LRO)
of vector chirality in the case of anisotropic exchange
couplings.\cite{NersesyanGE1998,KaburagiKH1999,HikiharaKK2001}

With applied magnetic field, the phase diagram becomes even richer.
From numerical studies of the magnetization process,
it has been found that for a certain range of $J_2/J_1$
the magnetization curve exhibits a plateau at one-third of
the saturated magnetization and cusp
singularities.\cite{OkunishiT2003,OkunishiHA1999,OkunishiT2003B,TonegawaOONK2004}
In this $1/3$-plateau phase, the ground state has
a magnetic LRO of {\it up-up-down} structure.
Furthermore, it was found that away from the $1/3$-plateau and
at $J_2/J_1\gtrsim1$,
the total magnetization $S^z_{\rm tot} = \sum_l s^z_l$
changes in units of $\Delta S^z_{\rm tot} = 2$,
indicating that two spins form a bound pair and flip simultaneously
as the field $h$ increases.\cite{OkunishiT2003,OkunishiT2003B}
These characteristic changes in the magnetization process give
accurate estimates of phase boundaries,
which divide the parameter space into several regions
(see Fig.~\ref{fig:phasediagram} below),
although the magnetization process alone cannot give much information
on the nature of each phase.

Another interesting feature of the $J_1$-$J_2$ zigzag ladder in magnetic field
is a field-induced LRO of the vector chirality,
\begin{equation}
\kappa_l^{(n)} = ( {\bm s}_l \times {\bm s}_{l+n} )^z.
\label{eq:vector-chirality}
\end{equation}
In zero field, the vector chiral LRO has been found
when and only when the system has an easy-plane
anisotropy.\cite{NersesyanGE1998,KaburagiKH1999,HikiharaKK2001,HikiharaKKT2000,Hikihara2002}
In this case, due to the anisotropy, symmetry of the system in spin space
is lowered from isotropic $SU(2)$ to $U(1) \times Z_2$,
where the $U(1)$ and $Z_2$ symmetries correspond to
the rotation in the easy plane and
the sign of pitch angle of helical spin order, respectively.
While the continuous $U(1)$ symmetry is preserved
in the quantum case $s = |{\bm s}| < \infty$,\cite{Momoi1996}
the discrete $Z_2$ symmetry can be spontaneously broken
even in the quantum limit $s=1/2$,
thereby resulting in the vector chiral phase.
This line of symmetry consideration suggests that
the magnetic field, which induces the same symmetry reduction,
should also lead to the spontaneous symmetry breaking of the $Z_2$ symmetry.
Indeed, this possibility was first pointed out
by Kolezhuk and Vekua,\cite{KolezhukV2005}
who have predicted from a field-theoretical analysis that
the vector chiral LRO may set in for a large $J_2/J_1$ regime.
Recently, the appearance of the vector chiral LRO under magnetic field
was verified numerically.\cite{McCullochKKKSK2008,Okunishi2008}

In this paper, we report our numerical and analytic results of
the ground-state properties in
the various phases that appear under magnetic field.
From a thorough comparison of long-distance
behavior of correlation functions,
we identify effective theories that describe
the low-energy physics of each phase.
For this purpose, we calculate numerically various correlation functions,
which include longitudinal-spin, transverse-spin, vector chiral, and
nematic (two-magnon) correlation functions
using the DMRG method.\cite{White1992,White1993}
Comparing the numerical results with asymptotic forms derived
from bosonization analysis,
we find that, in addition to the gapped dimer phase, $1/3$-plateau phase, and
fully polarized phase, the system exhibits four critical phases:
(i) a phase with one-component TLL which is adiabatically connected
to the ground state of the 1D Heisenberg antiferromagnet (TLL1 phase),
(ii) a two-component TLL phase (TLL2 phase),
(iii) a vector chiral phase,
and (iv) a spin-density-wave phase with two-spin bound pairs (SDW$_2$ phase).
The low-energy states in the TLL1, vector chiral, and SDW$_2$ phases
turn out to be one-component TLLs (a conformal field theory with
central charge $c=1$).
Furthermore, we provide quantitative estimates of non-universal parameters 
appearing in the low-energy effective theories, such as the TLL parameter 
and incommensurate wavenumbers of spin correlations, 
as functions of $J_2/J_1$ and the magnetization.
In particular, our results of the TLL parameter, which controls
decay exponents of correlation functions,
have direct relevance to experimental observables,
e.g., a magnetic LRO emerging in real quasi-1D compounds
with weak interladder couplings
and temperature dependence of relaxation rates ($1/T_1$)
in nuclear magnetic resonance experiments.\cite{Giamarchi-text,SatoMF}
We also propose a two-component TLL theory to describe the TLL2 phase.

This paper is organized as follows:
In Sec.\ \ref{sec:diagram}, we show the ground-state
phase diagram under magnetic field (see Fig.~\ref{fig:phasediagram}),
which contains the TLL1, $1/3$-plateau, SDW$_2$,
vector chiral, TLL2, dimer, and fully polarized phases.
We briefly summarize the characteristics of each phase.
In the following sections, we discuss in detail our numerical results
for correlation functions and effective theories for each phase.
In Sec.\ \ref{sec:TLL1}, we consider the TLL1 phase,
which appears in small $J_2/J_1$ regime.
The correlation functions obtained with the DMRG method are shown to
be fitted well to analytic forms obtained from a bosonization theory
for a weakly-perturbed single Heisenberg spin chain, and
the decay exponents
of the spin correlation functions are estimated accurately.
This analysis reveals that the dominant correlation function changes
from the staggered transverse-spin correlation
to incommensurate longitudinal-spin one as $J_2/J_1$ increases.
In Sec.\ \ref{sec:SDW2}, we discuss the SDW$_2$
phase, which appears at larger $J_2/J_1$.
From the fitting of numerical data to bosonization theory,
we show that the low-energy excitations
are described by a one-component TLL with quasi-long-ranged
dominant incommensurate longitudinal-spin
and subleading nematic correlations
and short-ranged transverse-spin correlation.
Section~\ref{sec:plateau} discusses
the 1/3-plateau phase.
We show that the numerically found up-up-down spin structure is understood
in terms of the bosonization theories for the neighboring
TLL1 and SDW$_2$ phases.
In Sec.\ \ref{sec:VC}, we consider the vector chiral phase,
which is also a one-component TLL.
The fitting analysis shows that the vector chiral phase is
characterized by the vector chiral LRO and
the incommensurate quasi-LRO of the transverse spins.
In Sec.\ \ref{sec:TLL2}, we develop a two-component TLL theory,
i.e., two free boson theories (central charge $c=1+1$),
as a low-energy effective theory for the TLL2 phase.
We confirm the central charge $c=2$ through numerical
computation of entanglement entropy.
The consistency between the effective theory and the DMRG result
is shown by examining a few dominant Fourier components in
the local magnetization profile near open boundaries.
Section~\ref{sec:conc} contains summary and discussions on
implications of our results to real quasi-1D compounds
with weak interladder couplings.

\section{Phase diagram}\label{sec:diagram}

\begin{figure}
\begin{center}
\includegraphics[width=65mm]{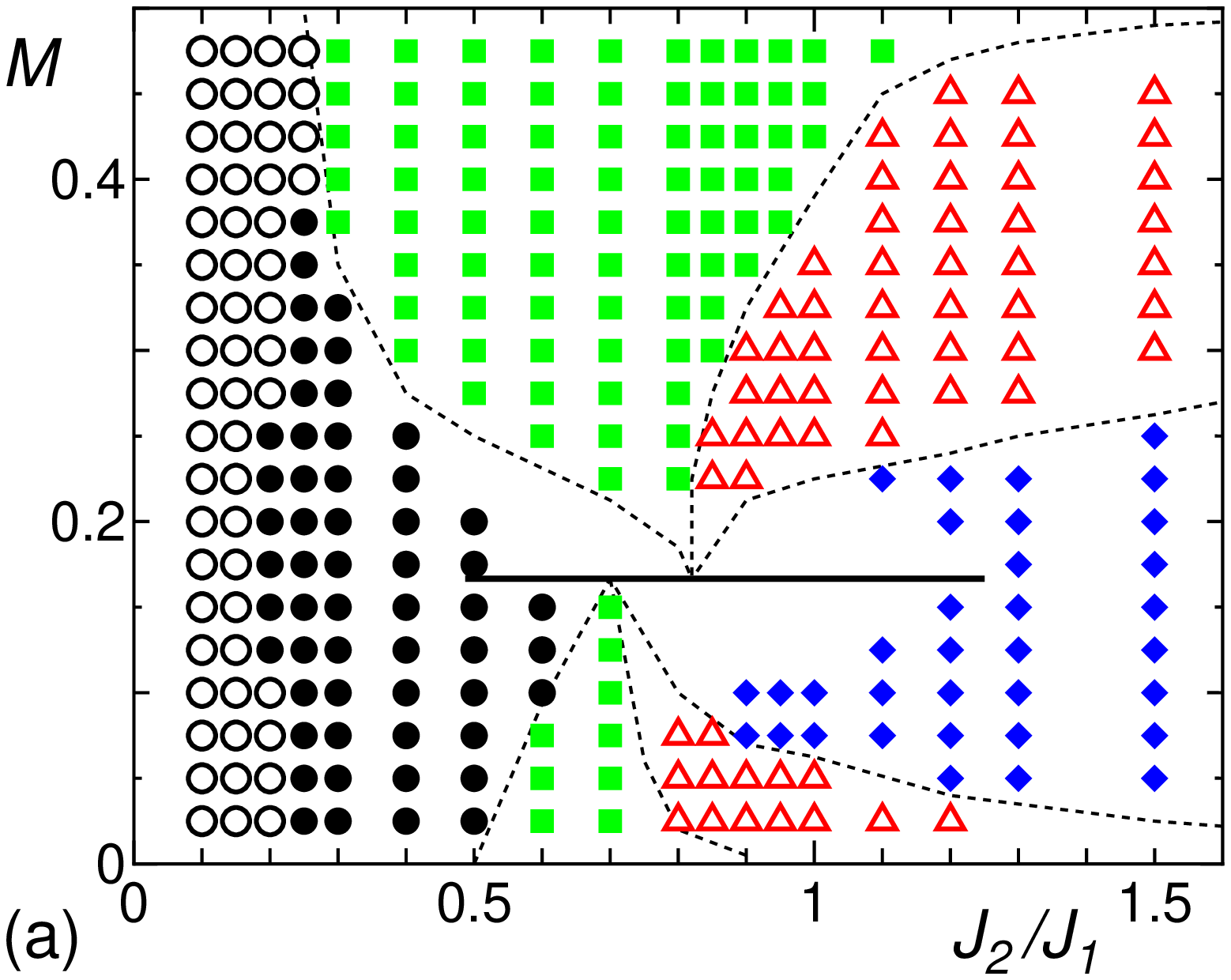}
\includegraphics[width=65mm]{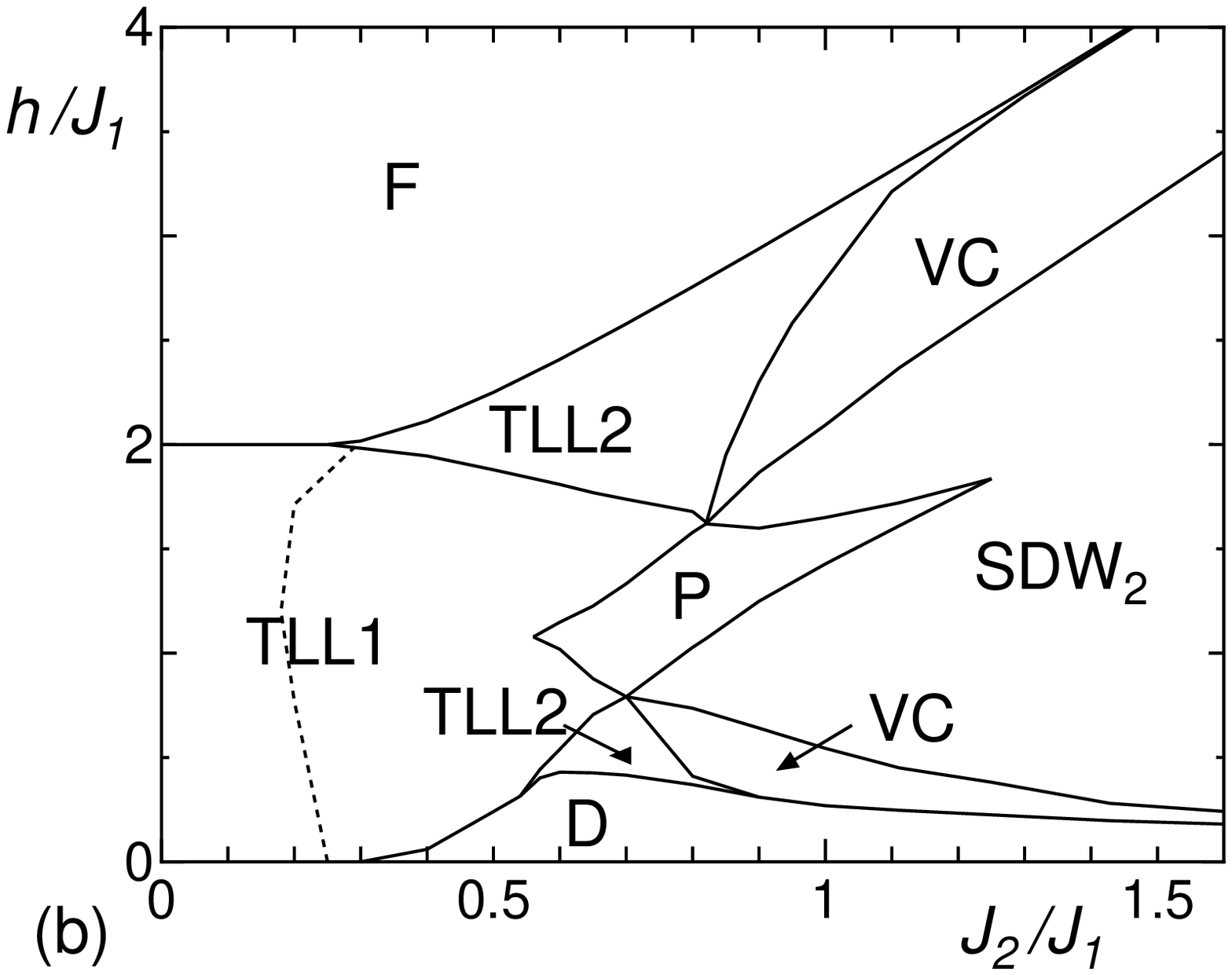}
\caption{
(Color online)
Magnetic phase diagram of the spin-1/2 antiferromagnetic zigzag ladder
(a) in the $J_2/J_1$ versus $M$ plane
and (b) in the $J_2/J_1$ versus $h/J_1$ plane.
In (a), symbols represent parameter points for which
their ground-state phases are identified:
Open ($\circ$) and solid ($\bullet$) circles represent the TLL1 phase
with dominant transverse- and longitudinal-spin correlation, respectively.
Diamonds ($\diamond$), triangles ($\vartriangle$), and squares ($\square$)
respectively represent
the SDW$_2$, vector chiral, and TLL2 phases.
The solid line shows the 1/3-plateau phase.
The dotted curves are the guide for the eye.
In (b), symbols P, VC, D, and F indicate
the 1/3 plateau, vector chiral, dimer, and
fully polarized phases, respectively.
The phase boundaries shown by solid lines are obtained
in Ref.~\onlinecite{OkunishiT2003}
from the numerical results of magnetization curves,
except for the boundaries between the vector chiral and TLL2 phases
which are obtained from the analysis of correlation functions
in the present paper.
The dotted line in the TLL1 phase represents the crossover line
between the transverse- and longitudinal-spin dominant regimes.
}
\label{fig:phasediagram}
\end{center}
\end{figure}

\begin{figure*}
\begin{center}
\includegraphics[width=160mm]{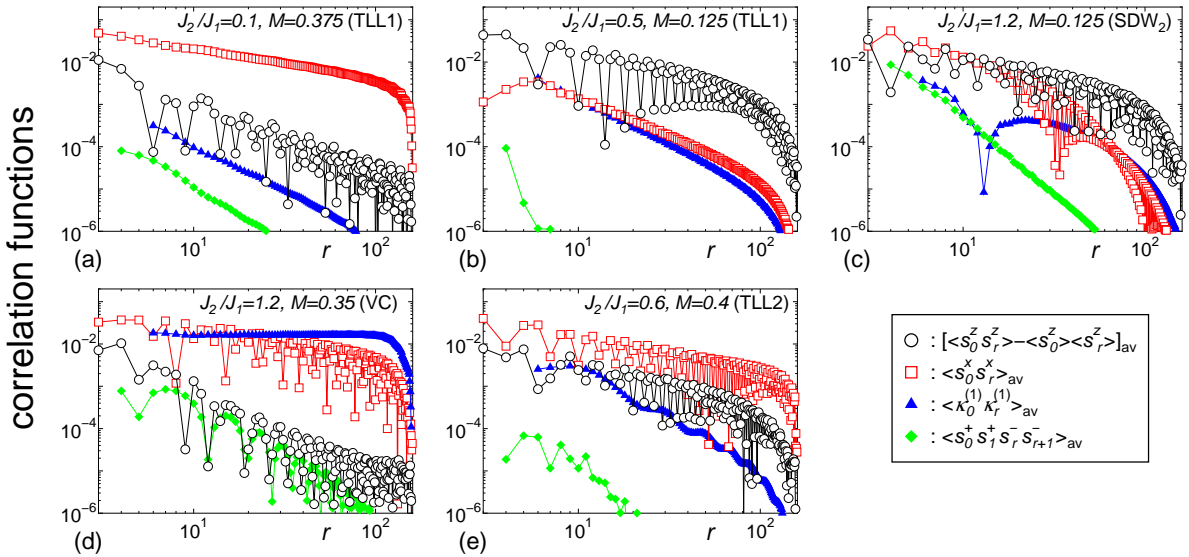}
\caption{
(Color online)
Typical spatial dependence of correlation functions in critical phases;
(a) TLL1 phase (transverse-spin correlation dominant),
(b) TLL1 phase (longitudinal-spin correlation dominant),
(c) SDW$_2$ phase,
(d) vector chiral phase, and
(e) TLL2 phase.
Absolute values of the averaged correlation functions
are plotted.
}
\label{fig:correlations}
\end{center}
\end{figure*}

Figure\ \ref{fig:phasediagram} presents the magnetic phase diagram
based on the numerical results obtained in this paper as well as
in previous studies.
The diagram is shown in the $J_2/J_1$ versus magnetization $M$ plane
in Fig.~\ref{fig:phasediagram}(a)
and in the $J_2/J_1$ versus $h/J_1$ plane
in Fig.~\ref{fig:phasediagram}(b),
where $M = (1/L) \sum_l s^z_l$ is the magnetization per site
and $L$ the system size.
The system exhibits at least four critical phases, i.e.,
TLL1, TLL2, vector chiral, and SDW$_2$ phases,
in addition to three gapped phases including
the dimer phase at $M = 0$, the 1/3-plateau phase ($M=1/6$),
and the fully polarized phase ($M=1/2$).

It has been revealed that the magnetization process of
the zigzag ladder (\ref{eq:Ham}) has remarkable
features:\cite{OkunishiT2003,OkunishiHA1999,OkunishiT2003B,TonegawaOONK2004}
for small $J_2/J_1$ ($0.25 \le J_2/J_1\lesssim 0.7$),
the magnetization curve has
at most two cusp singularities at higher and lower fields,
$h = h_{\rm c1}$ and $h_{\rm c2}$,
which correspond to boundaries between the TLL1 and TLL2 phases.
A magnetization plateau also appears at $M = 1/6$
for $0.487 < J_2/J_1 \lesssim 1.25$ and
$h_{\rm p1} < h < h_{\rm p2}$.\cite{OkunishiT2003,TonegawaOONK2004}
For large $J_2/J_1$, the magnetization process exhibits
two-spin flips with $\Delta S^z_{\rm tot} = 2$ in an intermediate field
region $h_{\rm m1} < h < h_{\rm m2}$.
See Figs. 2 and 3 in Ref.~\onlinecite{OkunishiT2003} for these results.
At zero magnetization, the ground state is gapless for $J_2/J_1<0.2411$ and
dimerized for $0.2411<J_2/J_1$.
The spin gap in the dimerized phase vanishes at
a critical field $h_\mathrm{d}$.
The ground state is fully polarized above
the saturation field $h_\mathrm{s}$.
The critical fields $h_{\rm c1}$, $h_{\rm c2}$, $h_{\rm p1}$, $h_{\rm p2}$,
$h_{\rm m1}$, $h_{\rm m2}$, $h_\mathrm{d}$, and $h_\mathrm{s}$ are plotted
in Fig.~\ref{fig:phasediagram}(b) with solid lines.

To reveal the nature of ground states in each region,
we have calculated several correlation functions, using the DMRG method,
for the system with up to $L=160$ spins with open boundaries.
We have kept typically $300$ block states in the calculation
(up to $400$ states for some cases),
and confirmed the convergence of the calculation
by checking the dependence of results on the number of kept states.
We have calculated
the longitudinal-spin correlation function
$\langle s^z_l s^z_{l'} \rangle$,
the transverse-spin correlation function $\langle s^x_l s^x_{l'} \rangle$,
the vector chiral correlation function
$\langle \kappa_l^{(n)} \kappa_{l'}^{(n')} \rangle$
with $n,n' = 1,2$,
the nematic (two-magnon) correlation function
$\langle s^+_l s^+_{l+1} s^-_{l'} s^-_{l'+1} \rangle$,
and the local spin polarization $\langle s^z_l \rangle$,
where $\langle \cdots \rangle$ denotes
the expectation value in the ground state.
To lessen the open-boundary effects,
we have computed the two-point correlation functions
for several pairs of $(l,l')$ with fixed distance $r=|l-l'|$
and taken their average for the estimate of the correlation
at the distance $r$.
In the following, we use the notation $\langle \cdots \rangle _{\rm av}$
for the averaged correlation functions.

Figure\ \ref{fig:correlations} shows
typical spatial dependence
of averaged correlation functions in the critical phases.
We note that the bending-down behaviors of
the averaged correlation functions seen for large distance
(e.g., $r \gtrsim 100$ for $L=160$)
are due to boundary effects
and should not be confused with intrinsic behaviors in the bulk.
Analyzing the long-distance behavior of correlation functions
in each parameter regime,
we have determined the low-energy effective theory for each phase.
The parameter points in the phase diagram at which
numerical results are explained successfully by the effective low-energy
theory of the corresponding phase
are shown with symbols in Fig.~\ref{fig:phasediagram}(a).
We summarize properties of each phase below.

\textit{TLL1 phase}:
In small $J_2/J_1$ regime, the ground state is adiabatically
connected to the one-component TLL of the antiferromagnetic Heisenberg chain
with only $J_1$ under magnetic field.
For relatively large $J_2/J_1$ ($0.25 \le J_2/J_1$)
the boundaries of the TLL1 phase are defined by the cusp singularities
in the magnetization curve.\cite{OkunishiHA1999,OkunishiT2003}
In this phase, both the longitudinal-spin fluctuation
$\langle s^z_0 s^z_r \rangle - \langle s^z_0 \rangle \langle s^z_r \rangle$
and transverse-spin correlation functions
$\langle s^x_0 s^x_r \rangle$ decay algebraically.
The former shows incommensurate oscillations
with a wavenumber $Q = \pi (1 \pm 2M)$,
while the latter is staggered, $Q = \pi$.
The numerical estimation of the decay exponents, shown in Sec.~\ref{sec:TLL1},
indicates that the dominant correlation function
changes from the staggered transverse-spin correlation
to incommensurate longitudinal-spin correlation as $J_2/J_1$ increases
[see Figs.\ \ref{fig:correlations} (a) and (b)].
The TLL1 phase is thus divided by the crossover line
into two regions of different dominant correlations, as shown
in Fig.~\ref{fig:phasediagram}.

\textit{SDW$_2$ phase}:
For large $J_2/J_1$,  there is a phase where the magnetization process
changes by the steps of $\Delta S^z_{\rm tot} = 2$.\cite{OkunishiT2003}
We show in Sec.\ \ref{sec:SDW2} that this phase is described
by a one-component TLL theory, which was originally derived from the
weakly-coupled AF Heisenberg chains in the limit
$J_2/J_1\gg1$.\cite{KolezhukV2005,HMeisnerHV2006,VekuaHMH2007,HikiharaKMF2008}
The phase is characterized by
the quasi-long-ranged longitudinal-spin and
nematic correlation functions,
$\langle s^z_0 s^z_r \rangle - \langle s^z_0 \rangle \langle s^z_r \rangle$
and $\langle s^+_0 s^+_1 s^-_r s^-_{r+1} \rangle$,
which are dual to each other,
and by the short-ranged transverse-spin correlation function
$\langle s^x_0 s^x_r \rangle$ reflecting a finite energy gap
to single-spin-flip excitations, as shown in Fig.~\ref{fig:correlations}(c).
The longitudinal-spin correlation is incommensurate
with the wavenumber $Q_2 = \pm \pi (1/2 + M)$.
Numerical analyses of correlation functions reveal that
the longitudinal-spin correlation function is dominant
in the whole parameter region of this phase.
We thus call this phase the SDW$_2$ phase.
We note that the same phase
has been
found in the zigzag ladder (\ref{eq:Ham}) with ferromagnetic $J_1$ and
AF $J_2$ as well.\cite{Chubukov1991,CabraHP2000,HMeisnerHV2006,KeckeMF2007,VekuaHMH2007,HikiharaKMF2008,SudanLL2008,LauchliSL2009}

\textit{1/3-plateau phase}:
At one third of the saturated magnetization, $M=1/6$,
there is a magnetization-plateau phase in the intermediate parameter region
$0.487 \lesssim J_2/J_1 \lesssim 1.25$.\cite{OkunishiT2003,TonegawaOONK2004}
This phase is characterized by a field-induced excitation gap
and a spontaneous breaking of translational symmetry
accompanied by a magnetic LRO of the up-up-down structure.
The ground state is three-fold degenerate.
As shown in Sec.\ \ref{sec:plateau},
all two-point correlation functions
exhibit exponential decay, in accordance with the fully-gapped nature
of the phase.

\textit{Vector Chiral phase}:
The vector chiral phase is characterized by
the LRO of the vector chirality $\kappa^{(n)}$ as well as quasi-LRO of
incommensurate transverse spins, which decays algebraically in space.
The discrete $Z_2$ symmetry corresponding to the parity about a bond center
is broken spontaneously and
the ground state is doubly degenerate in the thermodynamic limit.
This vector chiral state is a quantum counterpart of the classical
helical state.
Though the classical helical state appears in $1/4 <J_2/J_1$ for arbitrary
magnetization, the quantum vector chiral phase is
found only in two narrow regions separated by the SDW$_2$ and
1/3-plateau phases,\cite{McCullochKKKSK2008,Okunishi2008}
see Fig.\ \ref{fig:phasediagram}.
We show that the vector chiral phase is also described by
a one-component TLL theory
which can be formulated starting from the two weakly-coupled AF
Heisenberg chains for
$J_2/J_1\gg1$.\cite{NersesyanGE1998,KolezhukV2005}
The correlation functions in this phase will be discussed
in Sec.\ \ref{sec:VC}.

\textit{TLL2 phase}:
The TLL2 phase occupies two parameter regions adjacent to the
TLL1 phase and the vector chiral phase.
The TLL2 phase is described as two Gaussian conformal field theories
(central charge $c=1+1$),
or a two-component TLL,
having two flavors of free massless bosonic fields
as its low-energy excitations.
In the Jordan-Wigner fermion representation, fermions have two
separate Fermi seas, and the two bosonic fields represent particle-hole
excitations near the two sets of Fermi points.
In the TLL2 phase all correlation functions decay algebraically and
have incommensurate wave numbers which are linear functions of
the two Fermi momenta of Jordan-Wigner fermions.
We will discuss these properties and the low-energy effective theory
in Sec.\ \ref{sec:TLL2}.

\textit{Dimer phase}:
For $J_2/J_1>0.2411$ and at $M=0$, the ground state of the
$J_1$-$J_2$ AF Heisenberg zigzag spin ladder is
spontaneously dimerized.\cite{MajumdarG1969A,MajumdarG1969B,Haldane1982,WhiteA1996,JullienH1983,OkamotoN1992,Eggert1996}
The ground state is doubly degenerate in the thermodynamic limit,
and there is a gap to lowest excitation.

\textit{Fully polarized phase}:
When applied magnetic field is larger than the saturation field,
$h > h_{\rm s}$,
the ground state is
in the fully polarized phase with saturated magnetization $M = 1/2$.
As the field decreases, the fully-polarized ground state
is destabilized by softening of single-magnon excitations,
which have the dispersion,
\begin{equation}
\varepsilon_k=J_1 (\cos k-1)+J_2(\cos 2k -1)+h.
\label{magnon}
\end{equation}
When $J_2/J_1<1/4$, the magnon dispersion has a single minimum at $k=\pi$,
while, when $J_2/J_1>1/4$, there are two energy minima at
$k=\pm \arccos(-J_1/4J_2)$.
The saturation field $h_{\rm s}$ is given by
$h_{\rm s}/J_1 = 2$ for $J_2/J_1 < 1/4$ and
$h_{\rm s}/J_1 = 2 J_2/J_1 + 1 + J_1/(8J_2)$ for $J_2/J_1 > 1/4$.

\section{TLL1 phase}\label{sec:TLL1}
In this section, we discuss the TLL1 phase appearing for small $J_2/J_1$.
Since the parameter space of this phase includes the AF Heisenberg chain
with $J_2 = 0$,
we naturally expect that the TLL1 phase should
share the same properties with the single Heisenberg chain.
Here, we first briefly review the TLL theory
for the AF Heisenberg zigzag ladder with weak $J_2$ coupling.
We then compare the theory with the numerical results of correlation functions
for the zigzag ladder (\ref{eq:Ham}) with $J_2 > 0$.

It is well known that the low-energy properties of a single Heisenberg chain
under magnetic field ($|M|<\frac12$ and $J_2 = 0$)
is described as a TLL.\cite{Giamarchi-text}
Since the (leading) operator generated from weak $J_2$ coupling
is irrelevant in applied magnetic
field (and marginally irrelevant without magnetic field) in the
renormalization-group sense,
the low-energy effective theory for small $J_2/J_1$ is
adiabatically connected to the TLL theory of the single AF Heisenberg chain
($J_2=0$).
Hence the low-energy excitations in the TLL1 phase
are free massless bosons
governed by the Gaussian model,
\begin{equation}
\widetilde{\mathcal{H}}_0
= \frac{v}{2} \int dx \left[ K \left( \frac{d\theta}{dx} \right)^2
+ \frac{1}{K} \left( \frac{d\phi}{dx}\right)^2 \right],
\label{eq:Ham0-TLL1}
\end{equation}
where $(\phi,\theta)$ are bosonic fields satisfying
the equal-time commutation relation
$[\phi(x), \partial_y\theta(y)] = i\delta(x-y)$.
The TLL parameter $K$ is a function of $J_2/J_1$ and $M$.
We have taken the lattice spacing to be one and
identify the continuous coordinate $x$ with the site index $l$.
The spin velocity $v$ is of order $J_1$, except for
the saturation limit $M \to 1/2$, where $v \to 0$.
The spin operators ${\bm s}_l$ can be expressed in terms
of the bosonic fields as
\begin{eqnarray}
s^z_l \!\!&=&\!\!
M + \frac{1}{\sqrt{\pi}} \frac{d\phi(x)}{dx}
\nonumber \\
&&\!\!{}
- (-1)^l a \sin[2\pi M l + \sqrt{4\pi} \phi(x)] + \cdots,
\label{eq:szl-TLL1}\\
s^+_l \!\!&=&\!\!
(-1)^l b e^{i \sqrt{\pi} \theta(x)}
\nonumber \\
&&\!\!{}
+ b' e^{i \sqrt{\pi} \theta(x)}
\sin[2\pi Ml + \sqrt{4\pi} \phi(x)] + \cdots,
\label{eq:s+l-TLL1}
\end{eqnarray}
where $a$, $b$, and $b'$ are nonuniversal positive constants,
whose numerical values are known at
$J_2=0$.\cite{HikiharaF1998,HikiharaF2001,HikiharaF2004}
Equations (\ref{eq:Ham0-TLL1}), (\ref{eq:szl-TLL1}), and (\ref{eq:s+l-TLL1})
define the effective theory for the TLL1 phase,
with which
asymptotic forms of spin correlation functions are obtained as
\begin{eqnarray}
\langle s^z_0 s^z_r \rangle \!\!&=&\!\!
M^2 - \frac{\eta}{4\pi^2 r^2}
+ A^z_1 \frac{(-1)^r \cos(2\pi M r)}{|r|^\eta} + \cdots,
\quad
\label{eq:Csz-TLL1}\\
\langle s^x_0 s^x_r \rangle \!\!&=&\!\!
A^x_0 \frac{(-1)^r}{|r|^{1/\eta}}
- A^x_1 \frac{\cos(2\pi M r)}{|r|^{\eta+1/\eta}} + \cdots,
\label{eq:Csx-TLL1}
\end{eqnarray}
where $A^z_1 = a^2/2$, $A^x_0 = b^2/2$, $A^x_1 = b'^2/4$
(with appropriate short-distance regularization),
and the decay exponent
$\eta$ is related to the TLL parameter $K$ by
$\eta = 2K$.
Equations (\ref{eq:Csz-TLL1}) and (\ref{eq:Csx-TLL1}) tell us that
for $\eta > 1$ the staggered transverse-spin correlation
function $\langle s^x_l s^x_{l'} \rangle$ is dominant,
while the incommensurate longitudinal-spin correlation
$\langle s^z_l s^z_{l'} \rangle$ with a wavenumber $Q = \pi(1 \pm 2 M)$
is dominant for $\eta < 1$.
At $J_2 = 0$ the decay exponent $\eta$ can be calculated exactly
using Bethe ansatz;\cite{BogoliubovIK1986,CabraHP1998}
$\eta$ increases monotonically as $M$ increases,
from $\eta = 1$ at $M = 0$ to $\eta = 2$ for $M \to 1/2$.
Therefore, at $J_2 = 0$,
the transverse-spin correlation $\langle s^x_0 s^x_r \rangle$ is
always the most-slowly decaying one for $0 < M < 1/2$.
For finite $J_2>0$, the exact value of the exponent $\eta$ is known
in the limit $M \to 0$.
For $J_2/J_1 < 0.2411$
where the ground state at $M=0$ is in the TLL1 phase,
$\eta = 1$ at $M=0$ because of the SU(2) symmetry.
On the other hand, for $J_2/J_1 > 0.2411$, i.e.,
when the ground state at $M=0$ is
in the dimer phase,\cite{Schulz1980,ChitraG1997}
$\eta \to 1/2$ as $M\to0$.
This means that $\eta$ is singular at $J_2/J_1=(J_2/J_1)_c$
and $M=0$.

One can also derive the (same) effective theory for the TLL1 phase,
starting from the saturation limit for $0<J_2/J_1<1/4$.
In this limit the system can be viewed as a dilute gas of
interacting hard-core bosons (magnons) with one flavor,
as the magnon dispersion (\ref{magnon}) has a single minimum at $k=\pi$.
The hydrodynamic theory for the one-flavor interacting bosons
is nothing but the TLL theory, Eq.\ (\ref{eq:Ham0-TLL1}).\cite{Giamarchi-text}
This approach naturally gives the same asymptotic forms of spin correlators
as Eqs.~(\ref{eq:szl-TLL1}) and (\ref{eq:s+l-TLL1}).
Furthermore,
in the saturation limit $M \to 1/2$,
$\eta\to2$ in the TLL1 phase (i.e., $J_2/J_1 < 1/4$),
since the dilute limit of the hard-core bose gas is equivalent to
a free fermion gas.

Next we discuss our DMRG results
of the transverse and longitudinal spin correlation functions
$\langle s^x_l s^x_{l'}\rangle$ and $\langle s^z_l s^z_{l'}\rangle$
and the local spin polarization $\langle s^z_l \rangle$.
To achieve better numerical convergence and efficiency,
the DMRG calculation was done for finite systems ($L$ spins)
with open boundaries.
We thus compare the numerical results with the correlation functions
calculated analytically from the effective theory (\ref{eq:Ham0-TLL1})
by imposing appropriate boundary conditions
on the bosonic field $\phi$.\cite{HikiharaF1998,HikiharaF2001,HikiharaF2004}
To this end, we have taken the Dirichlet boundary conditions
$\phi(\delta)=\phi(L+1-\delta)=0$,\cite{Fath2003}
where $\delta$ is a free parameter to be determined later.
For example, spatial dependence of the magnetization is given by
\begin{equation}
\langle s^z_l \rangle
= z(l;q)
\equiv \frac{q}{2\pi}
- a \frac{(-1)^l \sin[q(l-\delta)]}
{f_{\eta/2}\biglb(2(l-\delta)\bigrb)},
\label{eq:szl-TLL1-finite}
\end{equation}
where
\begin{equation}
q = \frac{2\pi LM}{L+1-2\delta},
\label{eq:q-TLL1}
\end{equation}
and
\begin{equation}
f_\alpha(x) =
\left[ \frac{2(L+1-2\delta)}{\pi}
\sin\left(\frac{\pi |x|}{2(L+1-2\delta)}\right) \right]^\alpha.
\label{eq:f}
\end{equation}
In the limit $l \ll L$, Eq.\ (\ref{eq:szl-TLL1-finite})
reduces to
\begin{equation}
\langle s^z_l\rangle=
M
-\frac{(-1)^la}{[2(l-\delta)]^{\eta/2}}
 \sin[2\pi M(l-\delta)].
\label{szl-TLL1}
\end{equation}
The presence of open boundaries gives rise to ``Friedel oscillations''
in the local magnetization.
The wave number of the oscillations is ``$2k_F$'' of the
Jordan-Wigner fermions, which equals $Q = \pi(1 \pm 2M)$ for $L\gg1$.
Similarly, the longitudinal and transverse spin correlation functions
are modified by boundary contributions as
\begin{widetext}
\begin{eqnarray}
\langle s^z_l s^z_{l'} \rangle
\!\!&=&\!\!
Z(l,l';q)
\nonumber \\
&\equiv&\!\!
\left( \frac{q}{2\pi} \right)^2
- \frac{\eta}{4\pi^2}
\left[ \frac{1}{f_2(l-l')} + \frac{1}{f_2(l+l'-2\delta)} \right]
- \frac{qa}{2\pi}
\left[ \frac{(-1)^l \sin[q(l-\delta)]}
{f_{\eta/2}\biglb(2(l-\delta)\bigrb)}
+ \frac{(-1)^{l'} \sin[q(l'-\delta)]}
{f_{\eta/2}\biglb(2(l'-\delta)\bigrb)} \right]
\nonumber \\
&&\!\!{}
+
\frac{(-1)^{l-l'} a^2}
{2 f_{\eta/2}\biglb(2(l-\delta)\bigrb)
f_{\eta/2}\biglb(2(l'-\delta)\bigrb)}
\left\{ \cos[q(l-l')] \frac{f_\eta(l+l'-2\delta)}{f_\eta(l-l')}
- \cos[q(l+l'-2\delta)] \frac{f_\eta(l-l')}{f_\eta(l+l'-2\delta)} \right\}
\nonumber \\
&&- \frac{\eta a}{2\pi} \!
\left\{
  \frac{(-1)^l \cos[q(l-\delta)]}
{f_{\eta/2}\biglb(2(l-\delta)\bigrb)} [g(l+l'-2\delta) + g(l-l')]
+ \frac{(-1)^{l'} \cos[q(l'-\delta)]}
{f_{\eta/2}\biglb(2(l'-\delta)\bigrb)} [g(l+l'-2\delta) - g(l-l')]
\right\},
\label{eq:Csz-TLL1-finite}
\\
\langle s^z_l s^z_{l'} \rangle
\!\!&-&\!\! \langle s^z_l \rangle \langle s^z_{l'} \rangle
= Z(l,l';q) - z(l;q) z(l';q),
\label{eq:Csz-szsz-TLL1-finite} \\
\langle s^x_l s^x_{l'} \rangle
\!\!&=&\!\! X(l,l'; q)
\nonumber \\
&\equiv&\!\!
\frac{f_{1/2\eta}\biglb(2(l -\delta)\bigrb)
      f_{1/2\eta}\biglb(2(l'-\delta)\bigrb)}
{f_{1/\eta}(l-l') f_{1/\eta}(l+l'-2\delta)}
\left\{
\frac{(-1)^{l-l'} b^2}{2}
+ \frac{{\rm sgn}(l-l') b b'}{2}
\left[ \frac{(-1)^{l'}\cos[q(l -\delta)]}
            {f_{\eta/2}\biglb(2(l -\delta)\bigrb)}
     - \frac{(-1)^l \cos[q(l'-\delta)]}
            {f_{\eta/2}\biglb(2(l'-\delta)\bigrb)} \right]
\right. \nonumber \\
&& \!\!\left.
- \frac{b'^2}{4 f_{\eta/2}\biglb(2(l -\delta)\bigrb)
                 f_{\eta/2}\biglb(2(l'-\delta)\bigrb)}
\left[ \cos[q(l+l'-2\delta)]
       \frac{f_\eta(l-l')}{f_\eta(l+l'-2\delta)}
     + \cos[q(l-l')]
       \frac{f_\eta(l+l'-2\delta)}{f_\eta(l-l')} \right]
\right\},
\label{eq:Csx-TLL1-finite}
\end{eqnarray}
\end{widetext}
where
\begin{equation}
g(x) = \frac{\pi}{2(L+1-2\delta)}
\cot\!\left[ \frac{\pi x}{2(L+1-2\delta)} \right].
\label{eq:g}
\end{equation}
In the limit $|L/2-l|\ll L$ and $|L/2-l'|\ll L$, boundary effects go away,
and Eqs.\ (\ref{eq:Csz-TLL1-finite}) and (\ref{eq:Csx-TLL1-finite})
reduce to Eqs.\ (\ref{eq:Csz-TLL1}) and (\ref{eq:Csx-TLL1}).
In the fitting procedure discussed below,
we have optimized $\delta$ to achieve the best fitting
of $\langle s^z_l \rangle$ and $\langle s^z_l s^z_{l'} \rangle 
- \langle s^z_l \rangle \langle s^z_{l'} \rangle$,
whereas we set $\delta = 0$ for $\langle s^x_l s^x_{l'} \rangle$
as it has turned out that
the numerical data of
$\langle s^x_l s^x_{l'} \rangle$
can be fitted sufficiently well without optimizing $\delta$.

\begin{figure*}
\begin{center}
\includegraphics[width=160mm]{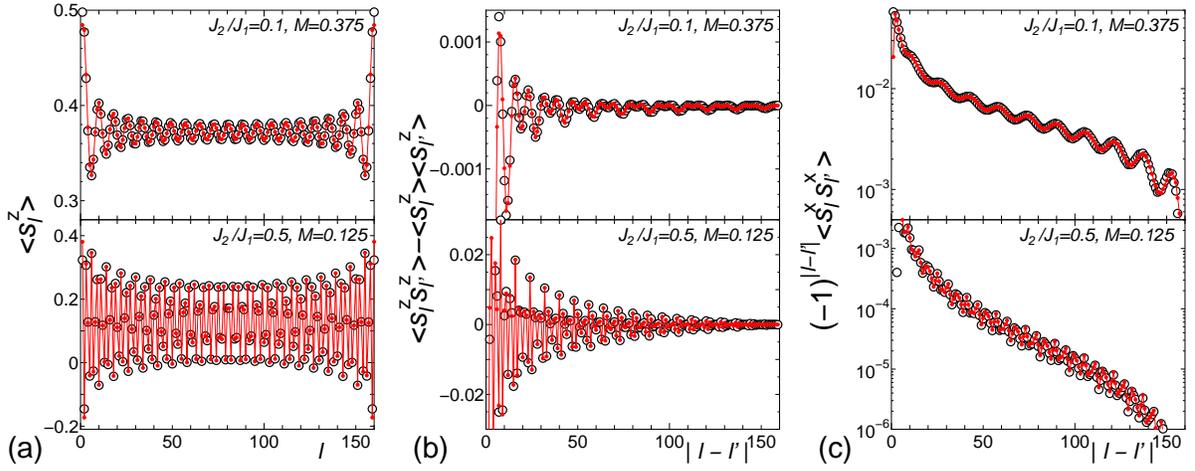}
\caption{
(Color online)
Correlation functions in the antiferromagnetic zigzag ladder
with $L=160$ spins in the TLL1 phase;
(a) local spin polarization $\langle s^z_l \rangle$,
(b) longitudinal-spin fluctuation
$\langle s^z_l s^z_{l'} \rangle
- \langle s^z_l \rangle \langle s^z_{l'} \rangle$,
and (c) transverse-spin correlation function $\langle s^x_l s^x_{l'} \rangle$.
The upper and lower panels show the results
for $(J_2/J_1, M) = (0.1, 0.375)$ and $(0.5, 0.125)$, respectively.
The open symbols represent the DMRG data
and the solid lines and circles are fits
to Eqs.\ (\ref{eq:szl-TLL1-finite}),
(\ref{eq:Csz-szsz-TLL1-finite}), and (\ref{eq:Csx-TLL1-finite}).
In (b) and (c), the data for $l = L/2 - [r/2]$ and $l' = L/2 + [(r+1)/2]$
are shown as a function of $r = |l-l'|$.
}
\label{fig:fit-TLL1}
\end{center}
\end{figure*}

Figure\ \ref{fig:fit-TLL1} shows
DMRG data of $\langle s^z_l \rangle$,
$\langle s^z_l s^z_{l'} \rangle
- \langle s^z_l \rangle \langle s^z_{l'} \rangle$,
and $\langle s^x_l s^x_{l'} \rangle$
for $(J_2/J_1, M) = (0.1, 0.375)$ and $(0.5, 0.125)$.
In the same figures, we show the fits to Eqs.\ (\ref{eq:szl-TLL1-finite}),
(\ref{eq:Csz-szsz-TLL1-finite}), and (\ref{eq:Csx-TLL1-finite}).
Clearly, the fits are in excellent agreement with
the numerical results.
We emphasize that only three fitting parameters,
$\eta$, $\delta$, and $a$ ($\eta$, $b$, and $b'$)
are used in the fitting of
 $\langle s^z_l \rangle$ and $\langle s^z_l s^z_{l'} \rangle
- \langle s^z_l \rangle \langle s^z_{l'} \rangle$
($\langle s^x_l s^x_{l'} \rangle$).
We have obtained almost the same good quality of fits
for the parameter points marked by open and solid circles
in Fig.~\ref{fig:phasediagram}(b),
which cover almost the entire region of the TLL1 phase.
These results thus demonstrate that the TLL1 phase
is described by the effective TLL theory given by
Eqs.\ (\ref{eq:Ham0-TLL1}), (\ref{eq:szl-TLL1}), and (\ref{eq:s+l-TLL1}),
which is indeed the same TLL theory as that of the AF Heisenberg
chain ($J_2=0$).

\begin{figure}
\begin{center}
\includegraphics[width=65mm]{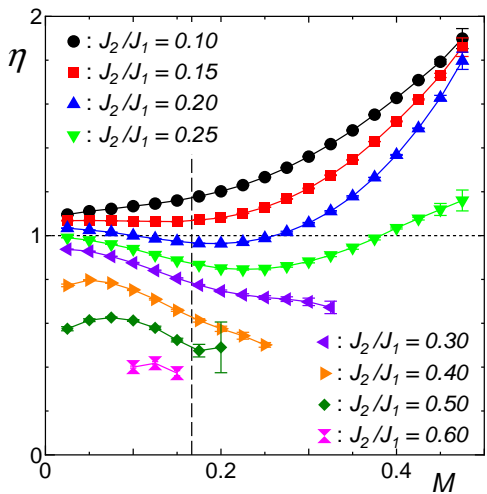}
\caption{
(Color online)
$M$ dependence of the exponent $\eta$ for the TLL1 phase
estimated from the fitting of $\langle s^x_l s^x_{l'} \rangle$
for the antiferromagnetic zigzag ladder with $L=160$ spins.
The error bars represent the difference of the estimates
obtained from the fitting of the data of different ranges.
The vertical dashed line corresponds to $M=1/6$
where the 1/3-plateau can appear for large $J_2/J_1$.
The dotted line at $\eta=1$ represents the boundary between 
the regimes of dominant staggered transverse-spin correlation ($\eta > 1$)
and dominant incommensurate longitudinal-spin correlation ($\eta < 1$); 
see Eqs.\ (\ref{eq:Csz-TLL1}) and (\ref{eq:Csx-TLL1}).
The exponent $\eta$ relates to the parameter $K$ as $\eta=2K$
in the TLL theory for the TLL1 phase.}
\label{fig:eta-M-TLL1}
\end{center}
\end{figure}

Figure\ \ref{fig:eta-M-TLL1} shows dependence of
the exponent $\eta$ on the magnetization $M$ in the TLL1 phase,
obtained from the fitting of the transverse-spin correlation
$\langle s^x_l s^x_{l'} \rangle$.
Similar estimates of $\eta$ are obtained from the other correlators
(not shown).
For small $J_2/J_1$, $\eta$ exhibits essentially the same behavior
as a function of $M$ as $\eta(M)$ at $J_2 = 0$;
for $J_2/J_1 \lesssim 0.15$, $\eta$ increases monotonically from
the universal value $\eta = 1$ at $M = 0$
to $\eta = 2$ at $M \to 1/2$ as $M$ increases.
In this regime the transverse-spin correlation
$\langle s^x_0 s^x_l \rangle$ is dominant for any $M$.
The situation changes as $J_2/J_1$ gets larger.
With increasing $J_2/J_1$, $\eta$ decreases
and becomes smaller than 1 at $J_2/J_1 = 0.2$
for intermediate magnetization $M$.
As $J_2/J_1$ is further increased in the TLL1 phase,
the exponent $\eta$ gets smaller than 1 for any $M$.
Thus, the system undergoes a crossover
from the small $J_2/J_1$ region with
the dominant staggered transverse-spin correlation
to the large $J_2/J_1$ region where the incommensurate
longitudinal-spin correlation
with $Q = \pi (1 \pm 2M)$ is dominant.
The crossover line is shown in the phase diagram,
Fig.~\ref{fig:phasediagram}.
The result is consistent with the earlier study,\cite{Gerhardt1997}
in which $\eta$ was estimated at $M=1/6, 1/4$ and $1/3$
for small systems.
Such a crossover between ground states with
the different dominant spin correlations has also been found
for the $J_1$-$J_2$ zigzag ladder with bond
alternation.\cite{UsamiS1998,HagaS2000,SuzukiS2004,MaeshimaOOS2004}

As mentioned above, $\eta$ is expected to approach
$1/2$ as $M \to 0$ for $J_2/J_1 > (J_2/J_1)_{\rm c} = 0.2411$.
Our numerical results at $J_2/J_1 = 0.5$ are consistent with
this theoretical prediction.
However, as $J_2/J_1$ approaches $(J_2/J_1)_{\rm c}$ from above,
the value of $\eta$ at the smallest $M = 0.025$ becomes larger
toward $\eta=1$, the value expected for $J_2/J_1 < (J_2/J_1)_\mathrm{c}$.
This implies that $\eta$ increases very rapidly from 1/2 at small $M$
for this parameter regime of $J_2/J_1 \lesssim 0.3$,
where the spin gap in the dimer ground state at $M=0$
is exponentially small (thereby small $M$ is sufficient to wipe out
dimer instability).

The data points for $0.3 \le J_2/J_1 \le 0.5$ end at the boundary
to the TLL2 phase for larger $M$.
Our results seem to indicate that $\eta$ changes continuously
along the TLL1-TLL2 phase boundary.

When the magnetization is close to $M=1/6$, the TLL1 phase
has an instability to the 1/3-plateau phase.
In the Jordan-Wigner fermion picture, the instability is caused
by umklapp scattering of three fermions, and the 1/3-plateau phase
corresponds to a density wave state of
the fermions.\cite{LecheminantO2004,HidaA2005}
The three-particle umklapp scattering is irrelevant at small $J_2/J_1$
but becomes relevant for larger $J_2/J_1$.
This explains why the 1/3-plateau phase emerges at $J_2/J_1 \gtrsim 0.5$
in the phase diagram (Fig.~\ref{fig:phasediagram}),
as we discuss below.

The effective Hamiltonian yielding the 1/3-plateau
has the form\cite{LecheminantO2004,HidaA2005,Notephi}
\begin{eqnarray}
\widetilde{\mathcal{H}} =
\widetilde{\mathcal{H}}_0
+ \lambda \int dx \sin\!\left[\pi(6M-1)x+3\sqrt{4\pi} \phi(x) \right],
\label{eq:Ham-plateau-TLL1}
\end{eqnarray}
where $\widetilde{\mathcal{H}}_0$ is
the Gaussian model (\ref{eq:Ham0-TLL1}) for the TLL1 phase
and $\lambda$ is the coupling constant for the three-particle
umklapp scattering.
The umklapp term
is accompanied by an oscillating factor with a wavenumber
$\pi(6M-1)\equiv 3\pi(1+2M)$
and becomes uniform at $M = 1/6$.
If we fix the magnetization at $M=1/6$ and increase $J_2/J_1$, then
the three-particle umklapp term becomes relevant for $K < 2/9$ ($\eta<4/9$).
Indeed, we see in Fig.~\ref{fig:eta-M-TLL1} that the estimates of
$\eta$ near $M = 1/6$ are larger than $4/9$ for $J_2/J_1\le0.4$
and become close to 4/9 at $J_2/J_1=0.5$.
This result is consistent with the estimated critical value
$(J_2/J_1)_{\rm p1} = 0.487$ which was obtained from the analysis of
the level spectroscopy in Ref.~\onlinecite{TonegawaOONK2004}.
For $J_2/J_1 > (J_2/J_1)_\mathrm{p1}$ we can approach the 1/3-plateau
phase by changing the magnetic field $h$.
This is in the universality class of commensurate-incommensurate
transition.\cite{Pokrovsky,Schulz1980}
In this case we expect that, as $M\to1/6$, the TLL parameter
$K$ approaches $1/9$, or, equivalently, $\eta\to2/9$.\cite{Schulz1980}
On the other hand, our numerical data for $J_2/J_1=0.6$ seem to be
much larger than the theoretical value $2/9$ at $M\to1/6$.
Although this disagreement might suggest that
there exist rather large errors in the estimates of $\eta$
for large $J_2/J_1$,
we rather expect that $\eta$ for $J_2/J_1 = 0.6$
should actually show rapid decrease very close to $M=1/6$
to recover the predicted behavior, $\eta \to 2/9$ as $M \to 1/6$.
Numerical verification of this would require calculations on much
larger systems.

\section{SDW$_2$ phase}\label{sec:SDW2}
In this section we discuss the SDW$_2$ phase.
This phase is characterized by two-spin flips $\Delta S^z = 2$
in the magnetization process.\cite{OkunishiT2003,EOphase}
The parameter space of the SDW$_2$ phase extends to large $J_2/J_1$,
see Fig.~\ref{fig:phasediagram}.
Its low-energy effective field theory is obtained
in the limit $J_2/J_1\to1$, and we will give
a short review on it
below.\cite{KolezhukV2005,HMeisnerHV2006,VekuaHMH2007,HikiharaKMF2008}
Then, by comparing our DMRG data of correlation functions
for $J_1/J_2\gtrsim1$ with the analytic results,
we demonstrate that the effective theory is valid
in the whole parameter space of the SDW$_2$ phase,
as expected from the principle of adiabatic continuity.\cite{BasicNotions}

In the limit $J_2 \gg J_1$, the zigzag spin ladder (\ref{eq:Ham})
can be viewed as two Heisenberg chains with {\it nearest-neighbor}
exchange $J_2$ coupled by weak interchain exchange $J_1$.
It is natural to bosonize each chain separately first
and then incorporate the interchain coupling $J_1$ perturbatively.
In this scheme, the original spin operators
are written as
\begin{align}
s^z_{2j+n} = & \,
M + \frac{1}{\sqrt{\pi}} \frac{d\phi_n(\bar{x}_n)}{d\bar{x}}
\nonumber \\
&
- (-1)^j a \sin[2\pi Mj + \sqrt{4\pi} \phi_n(\bar{x}_n)] + \cdots,
\label{eq:sz-boson-twochain} \\
s^+_{2j+n} = & \,
(-1)^j b \, e^{i \sqrt{\pi} \theta_n(\bar{x}_n)}
\nonumber \\
&
+ b' e^{i \sqrt{\pi} \theta_n(\bar{x}_n)}
\sin[2\pi Mj + \sqrt{4\pi} \phi_n(\bar{x}_n)] + \cdots,
\nonumber \\
\label{eq:s+-boson-twochain}
\end{align}
where $(\phi_n, \theta_n)$ are the bosonic fields for each chain $n=1,2$.
The coordinate $\bar{x}$ is related to the site index $l=2j+n$ ($n=1,2$) as
$\bar{x}_1 = j - 1/4$ and $\bar{x}_2 = j + 1/4$.
The low-energy theory of each AF Heisenberg chain has the same form
as $\widetilde{H}_0$ in Eq.\ (\ref{eq:Ham0-TLL1}).
The bosonized form of the interchain coupling $J_1$ can be found
from Eqs.\ (\ref{eq:sz-boson-twochain}) and (\ref{eq:s+-boson-twochain}).
We then obtain the effective Hamiltonian\cite{KolezhukV2005,HMeisnerHV2006,VekuaHMH2007,HikiharaKMF2008}
\begin{eqnarray}
\widetilde{\mathcal{H}} \!\!&=&\!\!
\sum_{\nu=\pm} \frac{v_\nu}{2} \int d\bar{x}
\left[
K_\nu \left( \frac{d\theta_\nu}{d\bar{x}} \right)^2
+
\frac{1}{K_\nu} \left( \frac{d\phi_\nu}{d\bar{x}} \right)^2
\right]
\nonumber \\
&&{}\!\!
+ g_1 \int d\bar{x} \sin(\sqrt{8\pi} \phi_- + \pi M)
\nonumber \\
&&\!\!{}
+ g_2 \int d\bar{x} \frac{d\theta_+}{d\bar{x}} \sin(\sqrt{2\pi} \theta_-),
\label{eq:tHam-two-chain}
\end{eqnarray}
where the interchain coupling gives the nonlinear interaction terms
with the coupling constants
\begin{equation}
g_1 = J_1 a^2 \sin(\pi M),~~~
g_2 = \frac{J_1}{2} \sqrt{2\pi} b^2
\end{equation}
in lowest order in $J_1$.
Here we have introduced symmetric $(+)$ and antisymmetric $(-)$ linear
combinations of the bosonic fields,
$\phi_\pm = (\phi_1 \pm \phi_2)/\sqrt2$,
$\theta_\pm = (\theta_1 \pm \theta_2)/\sqrt2$.
In lowest order in $J_1$ the TLL parameters $K_\pm$ are
given by
\begin{equation}
K_\pm = K \left( 1 \mp \frac{J_1 K}{\pi v}\right),
\label{eq:K+-}
\end{equation}
where $K$ is the TLL parameter of the decoupled Heisenberg
chains.\cite{KolezhukV2005}
This suggests that $K_+$ is less than 1
and decreases with $J_1$
at the limit $J_1 \ll J_2$.
The spin velocities $v_\pm$ are of order $J_2$ in the weak-coupling regime,
except for $M \to 1/2$ where $v_\pm \to 0$.

The effective Hamiltonian (\ref{eq:tHam-two-chain}) has
two competing interactions ($\propto g_1$ and $g_2$).
The fate of the ground state is determined by which one of
the two interactions grows faster in renormalization-group transformations.
If the $g_1$ term is dominant, the SDW$_2$ phase is realized.
We discuss this case below.
On the other hand, if the $g_2$ term is most relevant, then the ground state
is in the vector chiral phase; this case is discussed in Sec.\ \ref{sec:VC}.

Let us assume that the $g_1$ term wins the competition.
Then the field $\phi_-$ is pinned at a minimum of the potential
$g_1 \sin(\sqrt{8\pi} \phi_- + \pi M)$,
\begin{equation}
\langle \phi_- \rangle
= - \sqrt{\frac{\pi}{8}} \left( \frac{1}{2} + M \right),
\label{eq:pinned-SDW2}
\end{equation}
as $g_1\propto J_1 > 0$.
Since the pinned field $\phi_-$ can be taken as a constant,
the difference of (the uniform part of) two neighboring spins vanishes,
$s^z_{2j+1}-s^z_{2j+2} = \sqrt{2/\pi} \partial_{\bar{x}} \phi_- = 0$.
This means that the two spins are bound and explains
the steps $\Delta S^z_{\rm tot} = 2$ in the magnetization
process.\cite{HMeisnerHV2006}
The dual field $\theta_-$ fluctuates strongly
and we can therefore safely ignore the $g_2$ coupling.
The antisymmetric sector $(\phi_-, \theta_-)$ has
an energy gap, which corresponds to
the binding energy of the two-spin bound state.

Since the bosonic fields $(\phi_+, \theta_+)$
in the symmetric sector are not directly affected by the relevant
interchain couplings,
they remain gapless and constitute the one-component TLL.
The effective Hamiltonian for the SDW$_2$ phase is the Gaussian model,
\begin{equation}
\widetilde{\mathcal{H}}_+
= \frac{v_+}{2} \int d\bar{x}
\left[ K_+ \left( \frac{d\theta_+}{d\bar{x}} \right)^2
+ \frac{1}{K_+} \left( \frac{d\phi_+}{d\bar{x}}\right)^2 \right].
\label{eq:Ham0-sym}
\end{equation}
Equations\ (\ref{eq:sz-boson-twochain}), (\ref{eq:s+-boson-twochain}),
(\ref{eq:pinned-SDW2}), and (\ref{eq:Ham0-sym})
represent the TLL theory for the SDW$_2$ phase.

Straightforward calculations yield
the longitudinal-spin and nematic (two-magnon) correlation functions
in the thermodynamic limit,
\begin{eqnarray}
&&\langle s^z_0 s^z_r \rangle =
M^2 - \frac{\eta}{\pi^2 r^2} + \frac{\tilde{A}_1^z}{|r|^\eta}
\cos\left[ \pi r \left( \frac{1}{2} + M \right) \right] + \cdots,
\nonumber \\
\label{eq:Csz-SDW2} \\
&&\langle s^+_0 s^+_1 s^-_r s^-_{r+1} \rangle =
\frac{(-1)^r \tilde{A}_0}{|r|^{1/\eta}}
\nonumber \\
&&\hspace*{25mm}
- \frac{\tilde{A}_1(-1)^r}{|r|^{\eta+1/\eta}}
\cos\left[ \pi r \left( \frac{1}{2} + M \right) \right] + \cdots,
\nonumber \\
\label{eq:Cmp-SDW2}
\end{eqnarray}
where the exponent $\eta = K_+$, and we have introduced
positive numerical constants
$\tilde{A}_{0,1}$ and $\tilde{A}_1^z$.
These correlations are quasi-long-ranged and dual to each other.
If $\eta < 1$, the incommensurate SDW
correlation [the third term in Eq.\ (\ref{eq:Csz-SDW2})]
is the most dominant,
while the staggered nematic correlation is
the strongest for $\eta > 1$.
The perturbative result (\ref{eq:K+-}) indicates that
the incommensurate SDW correlation is dominant ($K_+<1$)
for small $J_1/J_2$.
We will see below that this holds true for $J_1/J_2\approx1$ as well.
The wavenumber of the SDW quasi-LRO is $Q_2 = \pm \pi (1/2 + M)$,
which is distinct from that of incommensurate correlations in other phases
and is characteristic of the SDW$_2$ phase.
We note that in the SDW$_2$ phase of the {\it ferromagnetic} ($J_1 < 0$)
$J_1$-$J_2$ zigzag ladder,
the characteristic wavenumber is
$Q= \pm \pi (1/2 - M)$.\cite{HikiharaKMF2008,SatoMF}
Such a spin-density-wave state with the incommensurate wave vector
is also found in the spatially-anisotropic triangular antiferromagnet
in magnetic field.\cite{StarykhB2007}

The transverse-spin correlation function
$\langle s^x_0 s^x_r \rangle$ decays exponentially
as the operator $s^x_l$ includes the strongly disordered $\theta_-$ field.
The exponential behavior is a direct consequence of the finite-energy cost
for creating a single-magnon excitation
and is a hallmark of the SDW$_2$ phase.

\begin{figure}
\begin{center}
\includegraphics[width=65mm]{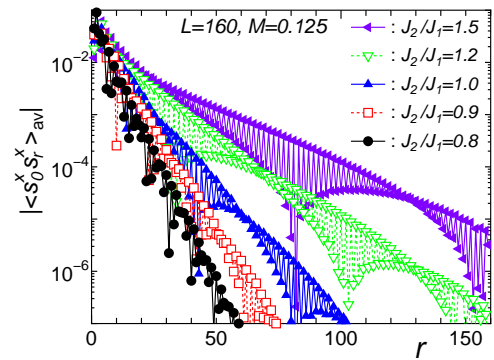}
\caption{
(Color online)
Absolute values of the averaged transverse-spin correlation function,
$|\langle s^x_0 s^x_r \rangle_{\rm av}|$,
in the antiferromagnetic zigzag ladder
with $L=160$ spins for $M = 0.125$ and several $J_2/J_1$.
}
\label{fig:cxx-SDW2}
\end{center}
\end{figure}

Let us discuss numerical results.
Figure\ \ref{fig:correlations}(c) shows typical behaviors
of the averaged correlation functions in the SDW$_2$ phase.
The longitudinal-spin and two-magnon correlation functions
decay algebraically and the former is clearly dominant.
By contrast,
as shown in Fig.~\ref{fig:cxx-SDW2},
the transverse-spin correlation decays exponentially.
This can be seen as evidence for the appearance
of two-magnon bound states in this parameter regime.
The correlation length of transverse spins becomes larger with
increasing $J_2/J_1$.
This is in accordance with the bosonization prediction that
the energy gap for the single-spin excitation is generated
by the cosine term with the coefficient $g_1 \sim J_1$ for $J_1/J_2\ll1$.
We have found essentially the same behavior of the correlation functions
as shown in Fig.~\ref{fig:correlations}(c)
for the entire parameter region where the two-spin-flips with
$\Delta S^z_{\rm tot} = 2$ are observed in the magnetization process.
After the dominant correlation function and the formation of
two-magnon bound pairs, we call this phase the SDW$_2$ phase.

\begin{figure}
\begin{center}
\includegraphics[width=65mm]{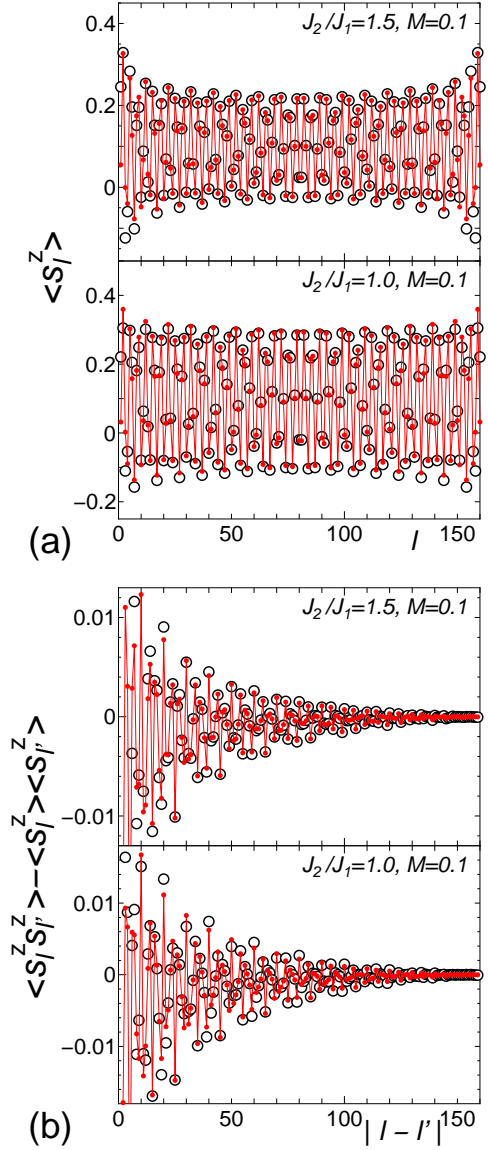}
\caption{
(Color online)
Correlation functions in the antiferromagnetic zigzag ladder
with $L=160$ spins in the SDW$_2$ phase;
(a) local spin polarization $\langle s^z_l \rangle$ and
(b) longitudinal-spin fluctuation
$\langle s^z_l s^z_{l'} \rangle
- \langle s^z_l \rangle \langle s^z_{l'} \rangle$.
The upper and lower panels show the results
for $(J_2/J_1, M) = (1.5, 0.1)$ and $(1.0, 0.1)$, respectively.
The open symbols represent the DMRG data
and the solid symbols are the results of fitting
to Eqs.\ (\ref{eq:szl-SDW2-finite}) and (\ref{eq:Csz-szsz-SDW2-finite}).
In (b), the data for $l = L/2 - [r/2]$ and $l' = L/2 + [(r+1)/2]$
are shown as a function of $r = |l-l'|$.
}
\label{fig:fit-SDW2}
\end{center}
\end{figure}

In order to estimate the exponent $\eta$
and to further demonstrate the validity of the effective theory
for the SDW$_2$ phase,
we fit the DMRG data of the local-spin polarization $\langle s^z_l \rangle$
and the longitudinal-spin fluctuation
$\langle s^z_l s^z_{l'} \rangle
- \langle s^z_l \rangle \langle s^z_{l'} \rangle$
to analytic forms obtained from the bosonization approach.
Using Eqs.\ (\ref{eq:sz-boson-twochain}), (\ref{eq:pinned-SDW2}),
and (\ref{eq:Ham0-sym}) and applying the Dirichlet boundary condition
in the same manner as in Sec.\ \ref{sec:TLL1},
we obtain the correlators for a finite open zigzag ladder as
\begin{align}
&\langle s^z_l \rangle = \frac{1}{2} - 2 z(l; \tilde{q}),
\label{eq:szl-SDW2-finite} \\
& \langle s^z_l s^z_{l'} \rangle
- \langle s^z_l \rangle \langle s^z_{l'} \rangle
= 4 \left[ Z(l,l';\tilde{q}) - z(l ;\tilde{q}) z(l' ;\tilde{q}) \right],
\label{eq:Csz-szsz-SDW2-finite}
\end{align}
with
\begin{equation}
\tilde{q} = \frac{2\pi L}{L+1-2\delta}
\left( \frac{1}{4} - \frac{M}{2} \right).
\end{equation}
In the limit $l\ll L$, Eq.\ (\ref{eq:szl-SDW2-finite}) reduces to
\begin{equation}
\langle s^z_l\rangle=
M
+\frac{2(-1)^la}{[2(l-\delta)]^{\eta/2}}
 \sin\left[\left(\frac{\pi}{2}-\pi M\right)(l-\delta)\right],
\label{Friedel-SDW2}
\end{equation}
showing Friedel oscillations with wavenumber $\pi(\frac{1}{2}+M)$.

Figure\ \ref{fig:fit-SDW2} shows DMRG results and their fits
to Eqs.\ (\ref{eq:szl-SDW2-finite}) and (\ref{eq:Csz-szsz-SDW2-finite}).
The results for $J_2/J_1 = 1.5$ show that the numerical data
at relatively large $J_2/J_1$ are fitted pretty well by the analytic forms.
Note that only three fitting parameters, $\eta$, $a$, and $\delta$,
are used in the fitting procedure.
For smaller $J_2/J_1$, the fitting results become less satisfactory,
presumably because
a smaller value of $\eta$ amplifies effects of both finite system size and
higher-order terms omitted in the analytic forms
(see also the discussion below for the estimate of $\eta$).
Nevertheless the fitting still gives a rather good result
at $J_2/J_1 = 1.0$ as well.
This observation gives a strong support to the validity
of the TLL theory for the SDW$_2$ phase.
We emphasize that the successful fitting directly demonstrates
that the characteristic wavenumber of the spin-density wave is
$Q_2 = \pm \pi (1/2 + M)$, in accordance with the theory above.
In the phase diagram in Fig.~\ref{fig:phasediagram}(a),
we plot the parameter points where the fitting worked well,
which cover almost the entire region of the SDW$_2$ phase.
In the vicinity of the 1/3-plateau, however, good fitting results
were not obtained due to the strong boundary effects.

\begin{figure}
\begin{center}
\includegraphics[width=65mm]{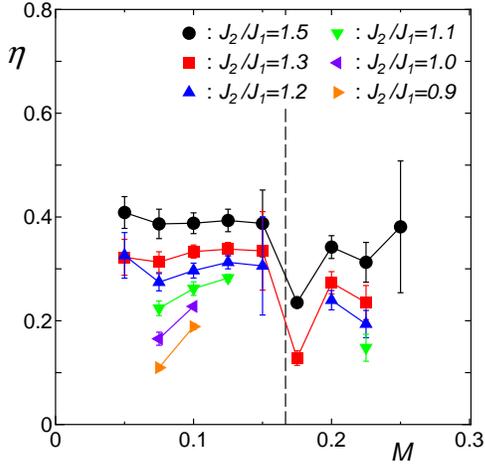}
\caption{
(Color online)
$M$ dependence of the exponent $\eta$ for the SDW$_2$ phase
estimated from the fitting of
$\langle s^z_l s^z_{l'} \rangle
- \langle s^z_l \rangle \langle s^z_{l'} \rangle$
for the antiferromagnetic zigzag ladder with $L=160$ spins.
The error bars represent the difference of the estimates
obtained from the fitting of the data of different ranges.
The vertical dashed line corresponds to $M=1/6$
where the 1/3-plateau can appear for small $J_2/J_1$.
The exponent $\eta$ relates to the parameter $K_+$ as $\eta=K_+$
in the TLL theory for the SDW$_2$ phase.
}
\label{fig:eta-M-SDW2}
\end{center}
\end{figure}

In Fig.~\ref{fig:eta-M-SDW2}, we present the exponent $\eta$
estimated from the fitting of the longitudinal-spin fluctuation
$\langle s^z_l s^z_{l'} \rangle
- \langle s^z_l \rangle \langle s^z_{l'} \rangle$.
Although the estimates have rather large error bars
coming from high sensitivity to the choice
of the data range used in the fitting,
we can safely conclude that the exponent for
$J_2/J_1 \lesssim 1.5$
is always small, i.e., $\eta \lesssim 0.5$.
This result reflects the fact that
the longitudinal-spin correlation is the strongest
in the SDW$_2$ phase.
Furthermore, the data show the tendency that
$\eta$ increases with $J_2/J_1$.
Combining this observation
with the perturbative result (\ref{eq:K+-}) for $J_2 \gg J_1$,
we may expect that $\eta$ increases monotonically with $J_2/J_1$
but less than 1 for the entire regime of $J_2/J_1 \gtrsim 1$.
This means that the SDW$_2$ phase, with the dominant
longitudinal-spin correlation, should extend from
the intermediate coupling regime of $J_2/J_1 \sim 1$
to the limit $J_2/J_1\to\infty$.

With decreasing $J_2/J_1$, the SDW$_2$ phase appears to touch
the 1/3-plateau phase.
Here we discuss this plateau-nonplateau transition
within the TLL theory for the SDW$_2$ phase.
We can consider the effective Hamiltonian with a three-particle
umklapp scattering,
\begin{eqnarray}
\widetilde{\mathcal{H}}'_+
=\widetilde{\mathcal{H}}_+
+ \tilde{\lambda} \int d\bar{x}
\sin\left[\pi\bar{x}(6M-1)+3 \sqrt{2\pi} \phi_+ \right],
\label{eq:Ham-plateau-SDW2}
\end{eqnarray}
where $\widetilde{\mathcal{H}}_+$ is given in Eq.\ (\ref{eq:Ham0-sym}),
and $\tilde{\lambda}$ is the coupling constant
for three-particle umklapp scattering.
The umklapp term
becomes uniform only at $M = 1/6$.
When $M=1/6$, the umklapp term is relevant for $K_+ < 4/9$.
Then the $\phi_+$ field is pinned and acquires a mass gap.
This results in the 1/3-plateau phase with up-up-down spin structure.
On the other hand, when approaching the 1/3-plateau
from incommensurate magnetization $M \to 1/6$,
$K_+$ takes the universal value $K_+ \to 2/9$.\cite{Schulz1980}
This is a commensurate-incommensurate transition.
Figure~\ref{fig:eta-M-SDW2} indicates that the estimated decay exponent
$\eta$ at slightly above $M = 1/6$
seems smaller than $4/9$ even for $J_2/J_1 = 1.5$,
suggesting the appearance of the 1/3-plateau at this coupling $J_2/J_1$.
This would mean that the upper critical value of the 1/3-plateau phase
is larger than 1.5, $(J_2/J_1)_{\rm p2} > 1.5$,
which is larger than the previous estimate $(J_2/J_1)_{\rm p2} \sim 1.25$
obtained from magnetization curves.\cite{OkunishiT2003}
While our estimated values of $\eta$ may contain some large errors,
another possible source of this discrepancy is that
the analysis of magnetization curves
could miss the plateau with an exponentially small width.
Further studies with higher accuracy will be needed for resolving this issue.

\section{1/3-plateau phase}\label{sec:plateau}

\begin{figure}
\begin{center}
\includegraphics[width=65mm]{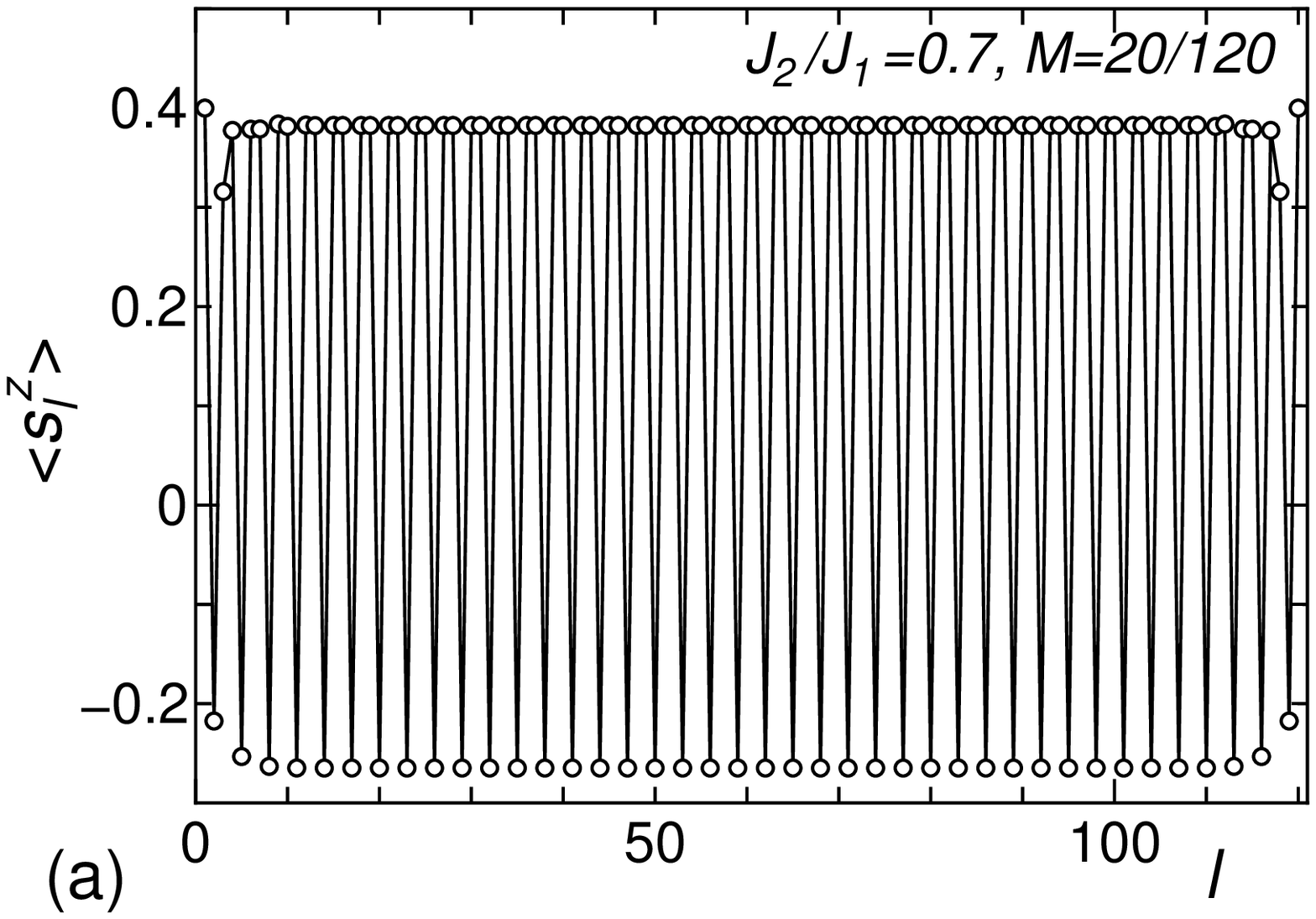}
\includegraphics[width=65mm]{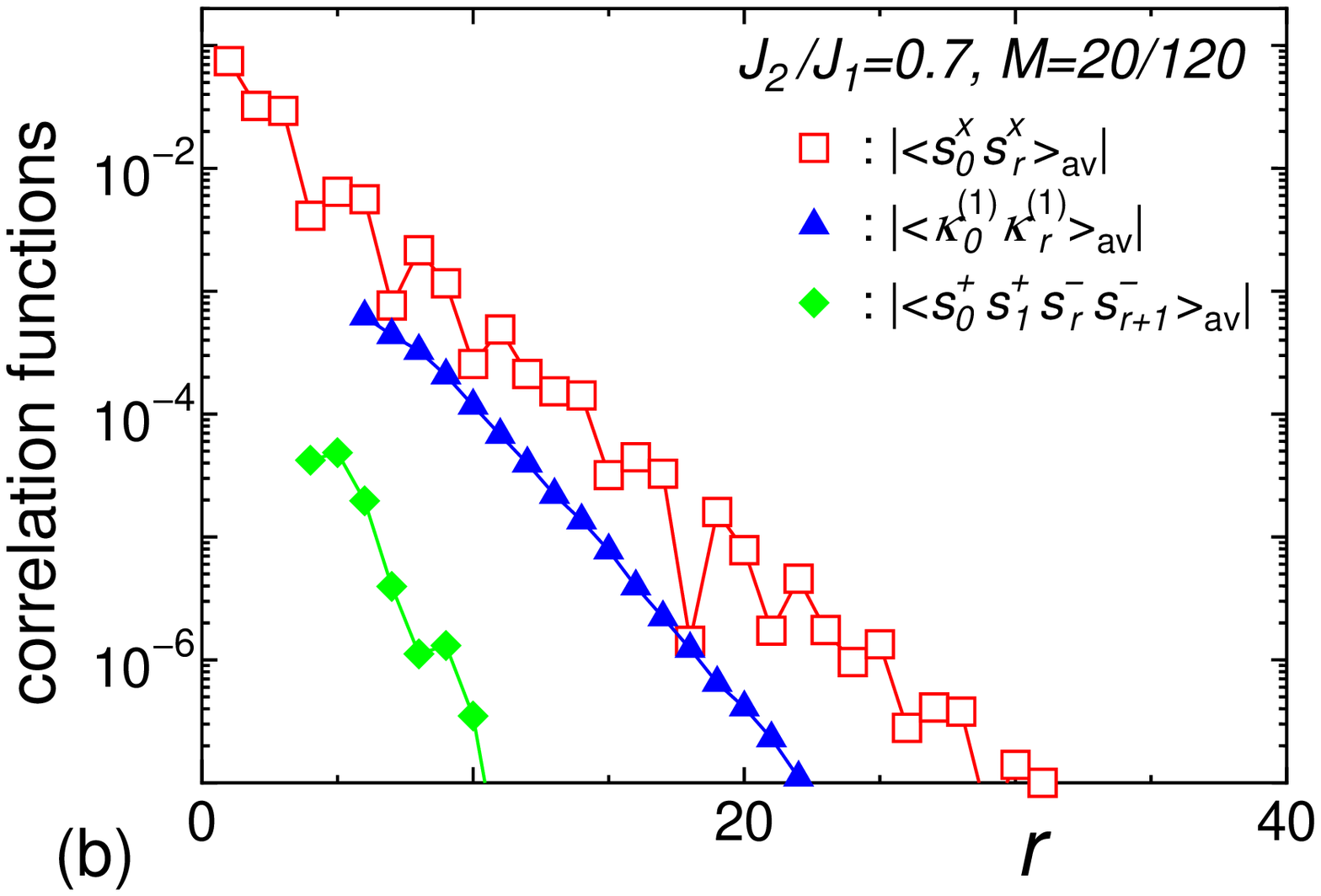}
\caption{
(Color online)
(a) Local magnetization $\langle s^z_l\rangle$ clearly shows
the up-up-down spin configuration.
(b) Semilog plot of the absolute values of the averaged correlation functions
in the antiferromagnetic zigzag ladder with $L=120$ spins
for $(J_2/J_1, M) = (0.7, 20/120)$.
}
\label{fig:cor-plateau}
\end{center}
\end{figure}

The 1/3-plateau phase with a finite spin gap
emerges at the magnetization $M = 1/6$
and for the parameter regime
$0.487 < J_2/J_1 \lesssim 1.25$.\cite{OkunishiT2003,OkunishiT2003B,TonegawaOONK2004}
In the 1/3-plateau phase the system has the magnetic LRO
of ``up-up-down" structure,\cite{OkunishiT2003,OkunishiT2003B}
as shown in Fig.~\ref{fig:cor-plateau}(a).
The ground state is therefore three-fold degenerate
in the thermodynamic limit.

The analysis of magnetization curves has shown that
the 1/3-plateau phase is surrounded by the TLL1 and SDW$_2$ phases
[see Fig.~1 of Ref.~\onlinecite{OkunishiT2003}
and Fig.~\ref{fig:phasediagram}(b) of the present paper].
As we discussed in Secs.~\ref{sec:TLL1} and \ref{sec:SDW2},
we can understand this phase diagram as the 1/3-plateau phase
emerging from instabilities of three-particle umklapp
scattering processes which are inherent
in the TLL1 and SDW$_2$ phases.
Here we shall discuss how the up-up-down spin configuration
emerges through pinning of bosonic fields.

When $J_2/J_1$ is small, the plateau emerges from the TLL1 phase.
As discussed in Sec.~\ref{sec:TLL1}, the transition is induced by
the three-particle umklapp scattering process.
If we fix the magnetization at $M=1/6$ and increase $J_2/J_1$,
the umklapp term becomes relevant at $K < 2/9$.
Indeed we observed
$K\simeq 2/9$ at $J_2/J_1=0.5$ in Fig.~\ref{fig:eta-M-TLL1},
which implies that for $J_2/J_1 \gtrsim 0.5$
the 1/3-plateau phase appears.
As the umklapp term is relevant,
the $\phi$ field is pinned at the bottom of the sine potential
in Eq.\ (\ref{eq:Ham-plateau-TLL1}),
$\sqrt{4\pi} \langle \phi \rangle = \pi/2$, $7\pi/6$, and $11\pi/6$
($\lambda>0$).
The bosonization formula of $s^z_l$, Eq.\ (\ref{eq:szl-TLL1}), then
reduces to
\begin{eqnarray}
s^z_l \!\!&=&\!\! \frac{1}{6}
-(-1)^la\sin\left(\frac{\pi l}{3}+\sqrt{4\pi}\langle\phi\rangle\right)
\nonumber\\
&=&\!\!\frac{1}{6}-a\cos\left(\frac{2\pi(l+n)}{3}\right),
\quad(n=0,1,2),
\label{eq:up-up-down1}
\end{eqnarray}
where $a>0$.
Equation (\ref{eq:up-up-down1}) gives
the up-up-down LRO with three-fold degeneracy
in the ground state.\cite{LecheminantO2004,HidaA2005,Notephi}

With larger $J_2/J_1$, the plateau phase is next to
the SDW$_2$ phase. As discussed in Sec.~\ref{sec:SDW2},
this phase transition is controlled by the three-particle umklapp
term, the second term in Eq.~(\ref{eq:Ham-plateau-SDW2}).
When $K_+ < 4/9$ and $M=1/6$, this term becomes relevant,
and the $\phi_+$ field is pinned to minimize the
potential energy.
The pinned values are
$\sqrt{2\pi} \langle \phi_+ \rangle = \pi/6$, $5\pi/6$, and $3\pi/2$
($\tilde\lambda<0$).
Substituting also $\phi_-=\langle\phi_-\rangle$ [Eq.\ (\ref{eq:pinned-SDW2})]
into the bosonized form of $s^z_l$,
Eq.\ (\ref{eq:sz-boson-twochain}),
yields
\begin{equation}
s^z_l=
\frac{1}{6}
+a\sin\left(\frac{2\pi l}{3}+\sqrt{2\pi}\langle\phi_+\rangle\right),
\end{equation}
which explains
the three-fold-degenerate ground state with the up-up-down LRO.

Since both $\phi_+$ and $\phi_-$ fields are pinned,
all low-energy excitations in the 1/3-plateau phase are gapped.
It thus follows that all correlation functions, except the long-ranged
longitudinal spin correlation, decay exponentially.
Figure\ \ref{fig:cor-plateau} shows 
the averaged correlation functions for $J_2/J_1 = 0.7$
and $M=1/6$ as a typical example for the 1/3-plateau phase.
The correlation functions decay exponentially
in accordance with the theory.

\section{Vector chiral phase}\label{sec:VC}
The vector chiral phase is characterized by
the spontaneous breaking of parity symmetry
accompanied by nonvanishing expectation value of the vector chirality,
$\langle \kappa_l^{(n)} \rangle
= \langle ({\bm s}_l \times {\bm s}_{l+n})^z \rangle \ne 0$.
The bosonization theory for the vector chiral phase
was developed in Refs.\ \onlinecite{NersesyanGE1998}
and \onlinecite{KolezhukV2005}, and
the appearance of the vector chiral LRO
in the zigzag spin ladder (\ref{eq:Ham})
has been numerically confirmed recently.\cite{McCullochKKKSK2008,Okunishi2008}
In this section we present results from our detailed numerical study
of correlation functions and compare them with their asymptotic forms
derived from the bosonization theory.

Let us first briefly summarize the results from the bosonization theory.
As discussed in Sec.\ \ref{sec:SDW2},
the effective Hamiltonian (\ref{eq:tHam-two-chain})
describes the zigzag spin ladder (\ref{eq:Ham})
in the limit $J_2 \gg J_1$.
When the $g_2$ term is most relevant,
we may employ the mean-field decoupling
approximation\cite{NersesyanGE1998} in which
both $d\theta_+/d\bar{x}$ and
$\sin(\sqrt{2\pi} \theta_-)$ are assumed to
acquire nonvanishing expectation values
to minimize the $g_2$ term.
The bosonic fields are thus pinned as
\begin{eqnarray}
\langle \theta_- \rangle = \mp \sqrt{\frac{\pi}{8}},~~~
\left\langle \frac{d\theta_+}{d\bar{x}} \right\rangle
 = \pm \sqrt{\frac{2}{\pi}} c,
\label{eq:pinned-VC}
\end{eqnarray}
where $c$ is a positive constant.
Selecting one set of the signs from $(+,-)$
and $(-,+)$ in Eq.\ (\ref{eq:pinned-VC})
corresponds to the spontaneous $Z_2$-symmetry breaking
in the vector chiral phase.
The antisymmetric sector $(\phi_-, \theta_-)$ thus acquires an energy gap
and the low-energy physics of the phase is governed by
the Gaussian model of the $(\phi_+, \theta_+)$ fields,
Eq.\ (\ref{eq:Ham0-sym}), in which the $\theta_+$ field has been redefined
as $\theta_+ \to \theta_+ - \langle d\theta_+/d\bar{x} \rangle \bar{x}$
to absorb the nonzero expectation value of
$\langle d\theta_+/d\bar{x} \rangle$.
The vector chiral phase is described by
a one-component TLL theory defined
by Eqs.\ (\ref{eq:sz-boson-twochain}), (\ref{eq:s+-boson-twochain}),
(\ref{eq:Ham0-sym}), and (\ref{eq:pinned-VC}).

Equation (\ref{eq:s+-boson-twochain}) allows us to write
the vector chiral operators $\kappa_l^{(n)}$ as
\begin{eqnarray}
\kappa_l^{(1)}
\!\!&\sim&\!\! \sin(\sqrt{2\pi} \theta_- ) ,
\label{eq:kappa1-VC} \\
\kappa_l^{(2)}
\!\!&\sim&\!\! \frac{d\theta_+}{d\bar{x}}.
\label{eq:kappa2-VC}
\end{eqnarray}
The nonvanishing expectation values in Eq.\ (\ref{eq:pinned-VC})
result in the vector chiral LRO in the ground state.
We note that
the expectation values of the vector chirality satisfy the relation
\begin{equation}
J_1\langle\kappa_l^{(1)}\rangle
+2J_2\langle\kappa_l^{(2)}\rangle
=0,
\label{eq:bloch-relation}
\end{equation}
so that there is no net spin current.\cite{HikiharaKMF2008}
Furthermore, one can easily obtain the leading asymptotic behaviors of
the transverse- and longitudinal-spin correlation functions as
follows:
\begin{eqnarray}
&& \langle s^x_0 s^x_r \rangle
= \frac{\tilde{A}^x}{|r|^{1/4K_+}} \cos(\widetilde{Q}r) +\cdots,
\label{eq:Csx-VC} \\
&& \langle s^x_0 s^y_r \rangle
= \pm \frac{\tilde{A}^x}{|r|^{1/4K_+}} \sin(\widetilde{Q}r) +\cdots,
\label{eq:Csxy-VC} \\
&&\langle s^z_0 s^z_r \rangle
= M^2 - \frac{K_+}{\pi^2 r^2} + \cdots,
\label{eq:Csz-VC}
\end{eqnarray}
where $\widetilde{Q} = (\pi+c)/2$, and $\tilde{A}^x$ is a
positive constant.
Equations\ (\ref{eq:Csx-VC}) and (\ref{eq:Csxy-VC}) indicate that
the spin components perpendicular to the applied field
have a spiral structure with the incommensurate wavenumber $\widetilde{Q}$,
which comes from the finite expectation value
of $\langle d\theta_+/d\bar{x} \rangle$.
This helical quasi-LRO of the transverse components
is a characteristic feature of the vector chiral phase.
The sign factor $\pm$ in Eq.\ (\ref{eq:Csxy-VC}) comes from
the sign $\pm$ in Eq.\ (\ref{eq:pinned-VC}),
and it defines the chirality, i.e., the direction of the spiral pitch.
In the longitudinal-spin correlation function,
the oscillating term with wavenumber $Q=\pi(\frac12+M)$
decays exponentially
as it includes the disordered $\phi_-$ field.
Therefore, if $1/(4K_+) < 2$,
the transverse-spin correlation function is dominant
except the long-ranged vector chiral correlations.

\begin{figure}
\begin{center}
\includegraphics[width=65mm]{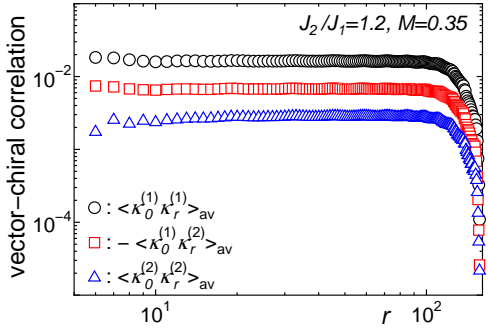}
\caption{
(Color online)
Averaged vector chiral correlation functions
in the antiferromagnetic zigzag ladder
with $L=160$ spins for $(J_2/J_1, M) = (1.2, 0.35)$.
Open circles, squares, and triangles respectively represent the DMRG data of
$\langle \kappa^{(1)}_0 \kappa^{(1)}_r \rangle_{\rm av}$,
$- \langle \kappa^{(1)}_0 \kappa^{(2)}_r \rangle_{\rm av}$, and
$\langle \kappa^{(2)}_0 \kappa^{(2)}_r \rangle_{\rm av}$,
where $\kappa^{(n)}_r = \left( {\bm s}_r \times {\bm s}_{r+n} \right)^z$.
}
\label{fig:vccor-VC}
\end{center}
\end{figure}

\begin{figure}
\begin{center}
\includegraphics[width=80mm]{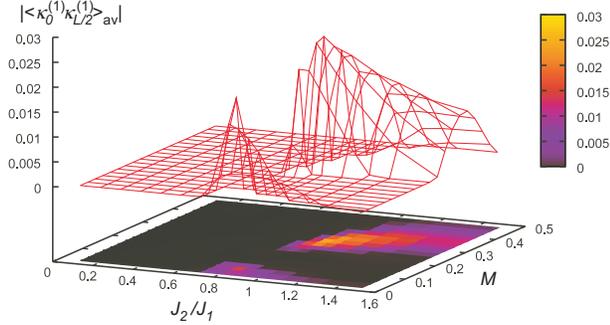}
\caption{
(Color online)
Amplitude of the vector chiral correlations at a distance $r = L/2$,
$|\langle \kappa^{(1)}_0 \kappa^{(1)}_{L/2} \rangle_{\rm av}|$,
in the antiferromagnetic zigzag ladder with $L=160$ spins.
}
\label{fig:VCLRO-VC}
\end{center}
\end{figure}

In Fig.~\ref{fig:vccor-VC}, we present our DMRG results of
the averaged vector chiral correlation functions
$\langle \kappa_0^{(n)} \kappa_r^{(n')} \rangle_{\rm av}$
for $(J_2/J_1, M) = (1.2, 0.35)$,
a representative point in the vector chiral phase.
Clearly, the vector chiral correlations are long-range ordered
(the reduction at $r>100$ are due to boundary effects
and should be ignored).
Figure\ \ref{fig:VCLRO-VC} shows $M$ and $J_2/J_1$ dependences of
the amplitude of the vector chiral correlations
measured at distance $r = L/2$,
$|\langle \kappa^{(1)}_0 \kappa^{(1)}_{L/2} \rangle_{\rm av}|$,
which indicates the strength of the LRO.
This figure shows the parameter regions of the vector chiral phase;
the parameter points where we observe the vector chiral LRO
are plotted in Fig.~\ref{fig:phasediagram}(b).
The vector chiral phase appears when $J_2/J_1$ is not small,
and the phase space is
split, by the SDW$_2$ phase, into two regions with
either small or large magnetization $M$.
This is in contrast with the $J_1$-$J_2$ zigzag ladder
with ferromagnetic $J_1$
and AF $J_2$ which has the vector chiral phase
only at small $M$.\cite{HikiharaKMF2008,SudanLL2008,LauchliSL2009}
It is also important to note that each one of the vector chiral phases
is next to a TLL2 phase [see Fig.~\ref{fig:phasediagram}(b)].
The amplitude of the vector chiral order parameter
exhibits a steep rise 
at the boundaries to the SDW$_2$ and TLL2 phases for small $J_2/J_1$ 
(see also Fig.\ 8 of Ref.\ \onlinecite{McCullochKKKSK2008}
for $J_2/J_1 = 1$) 
while the rise is modest for large $J_2/J_1$.
Incidentally, we have numerically confirmed that
the vector chiral correlations satisfy the relation (\ref{eq:bloch-relation}).
These observations on the vector chiral order
are consistent with the previous numerical 
results.\cite{McCullochKKKSK2008,Okunishi2008}

\begin{figure}
\begin{center}
\includegraphics[width=80mm]{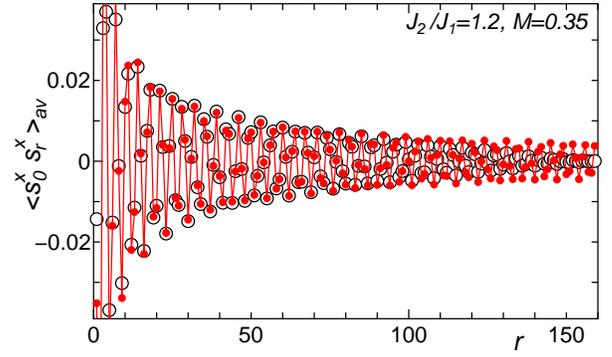}
\caption{
(Color online)
Averaged transverse-spin correlation function
$\langle s^x_0 s^x_r \rangle_{\rm av}$
in the antiferromagnetic zigzag ladder
with $L=160$ spins for $(J_2/J_1, M) = (1.2, 0.35)$.
Open circles represent the DMRG data.
The fits to Eq.\ (\ref{eq:Csx-VC}) are shown by solid circles.
}
\label{fig:cxx-VC}
\end{center}
\end{figure}

\begin{figure}
\begin{center}
\includegraphics[width=65mm]{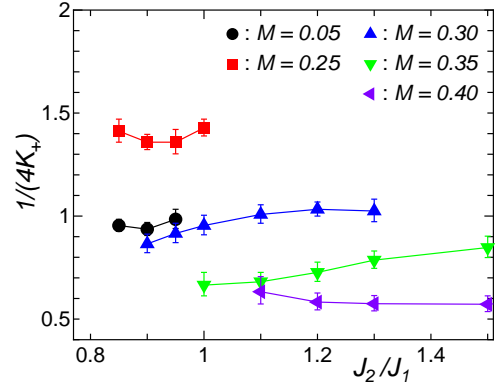}
\caption{
(Color online)
$J_2/J_1$ dependence of the decay exponent $1/(4K_+)$
of the transverse-spin correlation function
$\langle s^x_0 s^x_r \rangle_{\rm av}$ in the vector chiral phase
for the antiferromagnetic zigzag ladder with $L=160$ sites.
}
\label{fig:K+-VC}
\end{center}
\end{figure}

\begin{figure}
\begin{center}
\includegraphics[width=65mm]{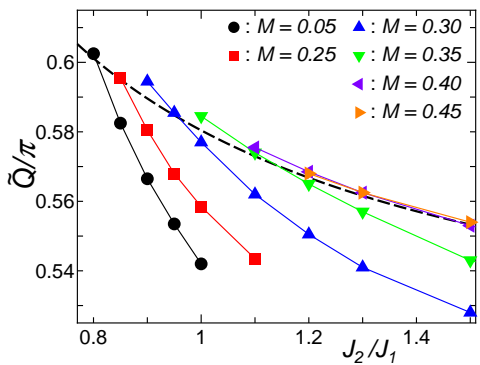}
\caption{
(Color online)
$J_2/J_1$ dependence of the wavenumber $\widetilde{Q}$
of the transverse-spin correlation function
$\langle s^x_0 s^x_r \rangle_{\rm av}$ in the vector chiral phase
for the antiferromagnetic zigzag ladder with $L=160$ sites.
The dashed curve represents the classical pitch angle
$\widetilde{Q} = \arccos(-J_1/4J_2)$.
}
\label{fig:Q-VC}
\end{center}
\end{figure}

To estimate the TLL parameter $K_+$ and the wavenumber $\tilde{Q}$
of the spiral transverse-spin correlation,
we fit the DMRG data of $\langle s^x_0 s^x_r \rangle_{\rm av}$
in the systems with $L=120$ and $160$ spins to Eq.\ (\ref{eq:Csx-VC}),
with taking $\widetilde{Q}$, $K_+$, and $\tilde{A}^x$ as fitting parameters.
Figure \ref{fig:cxx-VC} shows the result
for $(J_2/J_1, M) = (1.2, 0.35)$ and $L=160$.
We see that the DMRG data are fitted very well to the analytic form,
except for large distances $r \gtrsim 100$ where the boundary effect
is not negligible.
The good agreement between the numerical data and the fits
supports the validity of the TLL theory for the vector chiral phase.

The decay exponent $1/(4K_+)$ of the transverse-spin correlation
$\langle s^x_0 s^x_r \rangle_{\rm av}$ is shown in Fig.~\ref{fig:K+-VC}.
We have compared the estimates from $L=160$ and $120$ spins
and confirmed that the finite-size effect is negligible
in the data shown in the figure
but not so for some parameter points
(results for which are not shown in Fig.~\ref{fig:K+-VC})
in the very vicinity of the phase boundaries.
It turns out that the exponent is rather small, $1/(4K_+) \lesssim 1$,
in most parameter region of the vector chiral phase,
suggesting the dominant spiral transverse-spin correlation.
The exponent becomes larger, as we move closer to the 1/3-plateau phase.

Figure\ \ref{fig:Q-VC} shows the wavenumber $\widetilde{Q}$
of the transverse-spin correlation function.
While $\widetilde{Q}$ almost coincides
with the classical pitch angle $\arccos(-J_1/4J_2)$
near the boundary to the TLL2 phase,
it becomes smaller than the classical pitch angle with increasing $J_2/J_1$,
i.e., moving inside the vector chiral phase.
We thus find that the incommensurate wavenumber $\widetilde{Q}$
in the vector chiral phase is renormalized towards
the commensurate value $\pi/2$ due to quantum fluctuations.

\section{TLL2 phase}\label{sec:TLL2}

The TLL2 phase is a two-component TLL
consisting of two flavors of free bosons.
In this section, we develop its effective low-energy theory
based on the bosonization of Jordan-Wigner fermions.
We then discuss DMRG results, which support the effective theory.

The TLL2 phase is realized in two separated regions
of high and low magnetic fields in the magnetic phase diagram.
Here we first consider the high-field TLL2 phase,
for which the origin of the two bosonic modes can be easily understood
by examining the instability of the fully polarized phase.

Inside the fully polarized phase ($h > h_\mathrm{s}$),
the spin-wave excitation has a finite energy gap
and the dispersion relation is given by
Eq.\ (\ref{magnon}).
As the magnetic field is lowered, the energy gap decreases and
vanishes at the saturation field $h = h_\mathrm{s}$.
For $h<h_\mathrm{s}$, the soft magnons proliferate and collectively
form a TLL.
We notice that there are two distinct cases:

(i) When $J_2/J_1 < 1/4$, the bottom of the single-magnon dispersion is
at $k = \pi$ (mod $2\pi$).
Magnons with $k \approx \pi$ become soft and condense below
the saturation field $h_\mathrm{s}$,
yielding a one-component TLL.
Indeed, we have found the TLL1 phase in this case
(see Sec.\ \ref{sec:TLL1}).

(ii) When $J_2/J_1 > 1/4$,
the dispersion has two minima, $k = \pi \pm Q_0$
with $Q_0 = \arccos(J_1/4J_2)$.
Both magnons with $k=\pi+ Q_0$ and $\pi - Q_0$ become soft
and proliferate below the saturation field.
The resulting phase is the TLL2 phase which consists of
equal densities of two flavors of condensed magnons.
We note that, if the densities are not equal,
the vector chiral phase will be realized,\cite{KolezhukV2005}
as we will discuss later.

A similar argument should apply to the TLL2 phase appearing
at lower magnetic field.
The elementary excitation driving the instability of
the dimer ground state is a ``spinon,''
a domain wall separating two regions of different dimer
pattern.\cite{ShastryS1981,BrehmerKMN1998,OkunishiM2001}
For $J_2/J_1 < (J_2/J_1)_{\rm L}$, the dispersion of
the two-spinon state has a single minimum at $k = \pi$
and only one soft mode is relevant in destabilizing the dimer state.
The TLL1 phase is thus expected to show up for $M > 0$.
For $J_2/J_1 > (J_2/J_1)_{\rm L}$, on the other hand,
the two-spinon excitation spectrum exhibits a double-well structure
with minima at incommensurate momenta $k = \pm k_0$,
which leads to the TLL2 (or vector chiral) phase for $M > 0$.
The critical coupling at which
the lowest points deviate from $k=\pi$
has been estimated to be
$(J_2/J_1)_{\rm L} = 0.54$.\cite{OkunishiM2001}

\subsection{Two-component TLL theory}
\label{sec:two-component TLL theory}

In this subsection we describe the two-component TLL theory
of the high-field TLL2 phase in detail.
As we discussed above, this phase can be understood as a two-component
TLL emerging from condensation of two soft magnon modes.
This suggests to formulate a low-energy effective theory in terms
of interacting magnons.\cite{KolezhukV2005,UedaTotsuka2009}
Such an approach is valid and useful near the saturation field.
An alternative approach we adopt here is to formulate the low-energy theory
in terms of Jordan-Wigner fermions filling two separate Fermi seas.
Advantage of the latter approach is that it can be applied
in the whole TLL2 phase.
The connection to the magnon picture will also be discussed below.

\begin{figure}
\begin{center}
\includegraphics[width=70mm]{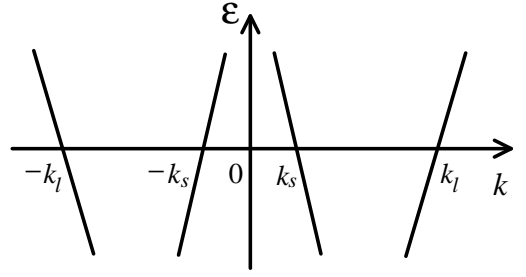}
\caption{
Four Fermi points located at $k=\pm k_s$ and $\pm k_l$.
The density of fermions is related to the magnetization,
$k_l-k_s=\pi(\frac{1}{2}-M)$.
The phase transition to the TLL1 phase occurs when $k_s=0$,
i.e., when the inner two Fermi points merge at $k=0$.
\label{fig:fermi}
}
\end{center}
\end{figure}

We apply the Jordan-Wigner transformation
\begin{subequations}
\begin{eqnarray}
s^z_l\!\!&=&\!\!
\frac{1}{2}-f^\dagger_lf^{}_l,
\label{JW-S^z_j}\\
s^+_l\!\!&=&\!\!
(-1)^lf^{}_l\exp\left(-i\pi\sum_{n<l}f^\dagger_nf^{}_n\right),
\label{JW-S^+_j}
\\
s^-_l\!\!&=&\!\!
(-1)^lf^\dagger_l\exp\left(i\pi\sum_{n<l}f^\dagger_nf^{}_n\right),
\label{JW-S^-_j}
\end{eqnarray}
\end{subequations}
to rewrite the Hamiltonian (\ref{eq:Ham}) in the form
$H=H_0+H'$, where
\begin{eqnarray}
H_0\!\!&=&\!\!
-\frac{J_1}{2}\sum_l
\left(f^\dagger_lf^{}_{l+1}+f^\dagger_{l+1}f^{}_l\right)
\nonumber\\
&&\!\!{}
+J_2M\sum_l
\left(f^\dagger_lf^{}_{l+2}+f^\dagger_{l+2}f^{}_l\right)
\nonumber\\
&&\!\!{}
-[2M(J_1+J_2)-h]\sum_lf^\dagger_lf^{}_l,
\label{H_0}
\end{eqnarray}
and
\begin{eqnarray}
H'\!\!&=&\!\!
J_1\sum_l:\!f^\dagger_lf^{}_l\!:\, :\!f^\dagger_{l+1}f^{}_{l+1}\!:
\nonumber\\
&&\!\!{}
+J_2\sum_l\left(
:\!f^\dagger_lf^{}_l\!:\,:\!f^\dagger_{l+2}f^{}_{l+2}\!:
-f^\dagger_{l+2}f^{}_l:\!f^\dagger_{l+1}f^{}_{l+1}\!:
\right.
\nonumber\\
&&\hspace*{14mm}\left.
-f^\dagger_lf^{}_{l+2}:\!f^\dagger_{l+1}f^{}_{l+1}\!:
\right).
\label{H'}
\end{eqnarray}
Here $:\!X\!:$ denotes normal ordering of $X$ with respect to
the filled Fermi sea of fermions with the dispersion
\begin{equation}
E(k)=-J_1\cos k+2J_2M\cos(2k)-2M(J_1+J_2)+h,
\label{dispersion}
\end{equation}
determined from Eq.~(\ref{H_0}).
Note that the wave number $k$ is measured from $\pi$
as the $(-1)^l$ factor is included in the Jordan-Wigner transformation.
As discussed above, in the TLL2 phase the dispersion has two minima
and, accordingly, there are four Fermi points located at $k=\pm k_s$,
$\pm k_l$ ($k_s<k_l$, see Fig.~\ref{fig:fermi}).
The density of fermions is
\begin{equation}
\rho=\frac{1}{\pi}(k_l-k_s)=\frac{1}{2}-M.
\label{density}
\end{equation}
In the limit $M\to\frac12$, both $k_l$ and $k_s$ approach $Q_0$.
Introducing slowly-varying fermionic fields for each Fermi point,
we write the fermion annihilation operator as
\begin{eqnarray}
f_j\!\!&=&\!\!
e^{ik_lx}\psi_{lR}(x)+e^{-ik_lx}\psi_{lL}(x)
\nonumber\\
&&\!\!{}
+e^{ik_sx}\psi_{sL}(x)+e^{-ik_sx}\psi_{sR}(x),
\label{slowly-varying fermionic fields}
\end{eqnarray}
where the continuous variable $x$ is identified with lattice index $j$.
We linearize the dispersion around the four Fermi points and
replace $H_0$ with
\begin{eqnarray}
\widetilde{H}_0\!\!&=&\!\!
iv_l\int dx\left(
\psi^\dagger_{lL}\frac{d}{dx}\psi^{}_{lL}
-
\psi^\dagger_{lR}\frac{d}{dx}\psi^{}_{lR}
\right)
\nonumber\\
&&\!\!\!{}
+iv_s\int dx\left(
\psi^\dagger_{sL}\frac{d}{dx}\psi^{}_{sL}
-
\psi^\dagger_{sR}\frac{d}{dx}\psi^{}_{sR}
\right),
\end{eqnarray}
where the velocities $v_l$ and $v_s$ are in general different.
The linearized kinetic term can be written as
\begin{equation}
\widetilde{H}_0=
\sum_{\nu=l,s}\frac{v_\nu}{4\pi}\int dx
\left[
\left(\frac{d\varphi_{\nu L}}{dx}\right)^2
+
\left(\frac{d\varphi_{\nu R}}{dx}\right)^2
\right]
\end{equation}
in terms of the chiral bosonic fields $\varphi_{\nu L}$ and $\varphi_{\nu R}$,
which obey the commutation relations
\begin{equation}
\begin{split}
&
[\varphi_{\nu R}(x),\varphi_{\nu R}(y)]
=i\pi\mathrm{sgn}(x-y),\\
&
[\varphi_{\nu L}(x),\varphi_{\nu L}(y)]
=-i\pi\mathrm{sgn}(x-y),\\
&
[\varphi_{\nu R}(x),\varphi_{\nu' L}(y)]
=-i\pi\delta_{\nu,\nu'}.
\end{split}
\end{equation}
The fermion densities are written as
\begin{equation}
\begin{split}
\rho_{\nu R}(x)&= \;
:\!\psi^\dagger_{\nu R}(x)\psi^{}_{\nu R}(x)\!: \;
=\frac{1}{2\pi}\frac{d\varphi_{\nu R}}{dx},
\\
\rho_{\nu L}(x)&= \;
:\!\psi^\dagger_{\nu L}(x)\psi^{}_{\nu L}(x)\!: \;
=\frac{1}{2\pi}\frac{d\varphi_{\nu L}}{dx}.
\end{split}
\end{equation}
Finally, the slowly-varying fermionic fields are bosonized,
\begin{equation}
\label{bosonization of fermion fields}
\begin{split}
\psi_{\nu R}(x)&=\frac{\eta_\nu}{\sqrt{2\pi\alpha}}e^{i\varphi_{\nu R}(x)},
\\
\psi_{\nu L}(x)&=\frac{\eta_\nu}{\sqrt{2\pi\alpha}}e^{-i\varphi_{\nu L}(x)},
\end{split}
\end{equation}
where $\alpha$ is a short-distance cutoff on the order of
the lattice spacing, and $\eta_\nu$ are the Klein factors
obeying $\{\eta_\nu,\eta_{\nu'}\}=2\delta_{\nu,\nu'}$.

The interaction Hamiltonian $H'$ gives rise to various scattering
processes of fermionic fields $\psi_{\nu L/R}$.
Among all, important in the TLL2 phase are
(short-range) density-density interactions,
\begin{eqnarray}
H_\rho\!\!&=&\!\!
\pi\!\int\!\!dx
\{
2g_{2l}\rho_{lL}(x)\rho_{lR}(x)
+
2g_{2s}\rho_{sL}(x)\rho_{sR}(x)
\nonumber\\
&&{}\hspace*{7mm}
+2g_{2\perp}[
\rho_{lL}(x)\rho_{sR}(x)+\rho_{lR}(x)\rho_{sL}(x)
]
\nonumber\\
&&{}\hspace*{7mm}
+
g_{4l}[\rho_{lL}(x)\rho_{lL}(x)+\rho_{lR}(x)\rho_{lR}(x)]
\nonumber\\
&&{}\hspace*{7mm}
+
g_{4s}[\rho_{sL}(x)\rho_{sL}(x)+\rho_{sR}(x)\rho_{sR}(x)]
\nonumber\\
&&{}\hspace*{7mm}
+2g_{4\perp}[
\rho_{lL}(x)\rho_{sL}(x)+\rho_{lR}(x)\rho_{sR}(x)
]
\},
\nonumber\\&&
\end{eqnarray}
where $g_{2l/s}$, $g_{2\perp}$, $g_{4l/s}$, and $g_{4\perp}$
are coupling constants that depend on $J_1$, $J_2$, and $M$.
We define the phase fields ($\nu=l,s$)
\begin{equation}
\begin{split}
&
\phi_\nu(x)=
\frac{1}{\sqrt{4\pi}}[\varphi_{\nu L}(x)+\varphi_{\nu R}(x)],
\\
&
\theta_\nu(x)=
\frac{1}{\sqrt{4\pi}}[\varphi_{\nu L}(x)-\varphi_{\nu R}(x)].
\end{split}
\end{equation}
The effective Hamiltonian $H_2=\widetilde{H}_0+H_\rho$
is then quadratic in $\phi_\nu$ and $\theta_\nu$,
and is diagonalized as
\begin{equation}
H_2=\int dx\sum_{\mu=\pm}\frac{v_\mu}{2}
\left[
\left(\frac{d\theta_\mu}{dx}\right)^2
+\left(\frac{d\phi_\mu}{dx}\right)^2
\right]
\label{eq:H_2}
\end{equation}
by the new fields $\theta_\pm$ and $\phi_\pm$ which are
linearly related to $\theta_{l,s}$ and $\phi_{l,s}$ by
\begin{equation}
\begin{pmatrix}
\phi_l \\ \phi_s
\end{pmatrix}
=A^T
\begin{pmatrix}
\phi_+ \\ \phi_-
\end{pmatrix},
\qquad
\begin{pmatrix}
\theta_l \\ \theta_s
\end{pmatrix}
=A^{-1}
\begin{pmatrix}
\theta_+ \\ \theta_-
\end{pmatrix}.
\label{A}
\end{equation}
Here the $2\times2$ matrix
\begin{equation}
A=
\begin{pmatrix}
A_{11} & A_{12} \\
A_{21} & A_{22} \\
\end{pmatrix}
\label{eq:A}
\end{equation}
is a function of the velocities $v_{l,s}$ and
the coupling constants $g$'s, whose functional form
can be found in Ref.~\onlinecite{Hikihara05}.
Without loss of generality, we can assume $v_+>v_-$.

The Hamiltonian $H_2$ (\ref{eq:H_2})
is the low-energy effective theory of the TLL2 phase.
It consists of two free bosonic sectors $(\phi_+,\theta_+)$
and $(\phi_-,\theta_-)$.
Other interactions which are not included in $H_\rho$
are irrelevant perturbations to $H_2$ in the TLL2 phase.
An important example of such interactions
is the backward-scattering interaction
\begin{eqnarray}
H_b\!\!&=&\!\!
g_{1\perp}\int dx
\left[
\psi^\dagger_{sL}(x)\psi^\dagger_{sR}(x)\psi^{}_{lL}(x)\psi^{}_{lR}(x)
+\mathrm{H.c.}
\right]
\nonumber\\
&=&\!\!
-\frac{g_{1\perp}}{2\pi^2\alpha^2}\int dx
\cos[\sqrt{4\pi}(\theta_l-\theta_s)].
\end{eqnarray}
The irrelevance of the operator
$\cos[\sqrt{4\pi}(\theta_l-\theta_s)]$ imposes the condition
\begin{equation}
\frac{1}{(\mathrm{det}\,A)^2}
[(A_{11}+A_{12})^2+(A_{21}+A_{22})^2]>2.
\end{equation}
We note that the vertex operators
$\exp(\pm i\sqrt{4\pi}\phi_\pm)$ and $\exp(\pm i\sqrt{4\pi}\theta_\pm)$
have scaling dimension 1.

The matrix $A$ takes a simple form
\begin{equation}
A=\frac{1}{\sqrt2}
\begin{pmatrix}
\sqrt{K_+} & 0 \\
0 & \sqrt{K_-}
\end{pmatrix}
\begin{pmatrix}
1 & 1 \\
1 & -1
\end{pmatrix},
\end{equation}
when the two conditions
\begin{subequations}
\label{conditions}
\begin{eqnarray}
&&
v_l+g_{4l}=v_s+g_{4s}\; =: v+g_4,
\\
&&
g_{2l}=g_{2s}\; =: g_2,
\end{eqnarray}
\end{subequations}
are satisfied.
In this case, the TLL parameters $K_\pm$ and
the renormalized velocities $v_\pm$ are given by
\begin{subequations}
\label{simple K and v}
\begin{eqnarray}
&&
K_\pm=
\left(
\frac{v+g_4\pm g_{4\perp}-g_2\mp g_{2\perp}}
     {v+g_4\pm g_{4\perp}+g_2\pm g_{2\perp}}
\right)^{1/2},
\\
&&
v_\pm=\left[
(v+g_4\pm g_{4\perp})^2-(g_2\pm g_{2\perp})^2
\right]^{1/2}.
\end{eqnarray}
\end{subequations}

This simplified effective theory is applicable
when $J_2/J_1\gg1/4$ and $|M-\frac12|\ll\frac12$,
i.e., when the magnon density
is very low and $k_l-k_s\ll k_s$.
In this case one can build an effective theory
by treating magnons with $k=\pi\pm Q_0$
as interacting hard-core bosons.\cite{KolezhukV2005,UedaTotsuka2009}
We adopt a phenomenological effective Hamiltonian
of interacting bosons ($0<v<u$),\cite{Cazalilla}
\begin{eqnarray}
H_\mathrm{B}\!\!\!&=&\!\!\!
\int dx\biggl[
\frac{1}{2m}\biggl(
\frac{d\psi_+^\dagger}{dx}\frac{d\psi^{}_+}{dx}
+
\frac{d\psi_-^\dagger}{dx}\frac{d\psi^{}_-}{dx}
\biggr)
\nonumber\\
&&
\hspace*{5mm}{}
+u\bigl\{[\rho_+(x)]^2+[\rho_-(x)]^2\bigr\}
+2v\rho_+(x)\rho_-(x)
\biggr],
\nonumber\\&&
\label{BosonHamiltonian}
\end{eqnarray}
where $\psi^{}_\pm(x)$ are field operators of two flavors of magnons
satisfying
$[\psi^{}_\mu(x),\psi^\dagger_{\mu'}(y)]=\delta_{\mu,\mu'}\delta(x-y)$,
magnon density fluctuations
$\rho_\pm(x)=\psi^\dagger_\pm(x)\psi^{}_\pm(x)-\rho/2$,
and $m$ is their effective mass.
The boson density (per flavor) is assumed to be $\rho/2$,
where $\rho$ is defined in Eq.\ (\ref{density}).
In the low-energy, hydrodynamic limit,\cite{Haldane1981,Giamarchi-text}
the magnon fields and density fluctuations are written as
\begin{subequations}
\label{hydro}
\begin{eqnarray}
&&
\psi_\pm^{}(x)\sim\sqrt{\frac{\rho}{2}}e^{i\vartheta_\pm(x)}+\ldots,
\\
&&
\rho_\pm(x)\sim
\frac{1}{\pi}\frac{d\varphi_\pm(x)}{dx}+
\rho\cos[\pi\rho x \! + \! 2\varphi_\pm(x)]+\ldots,  \qquad
\end{eqnarray}
\end{subequations}
where the phase fields obey
$[\varphi_\mu(x),\partial_y\vartheta_{\mu'}(y)]
=i\pi\delta_{\mu,\mu'}\delta(x-y)$.
Substituting (\ref{hydro}) into (\ref{BosonHamiltonian})
yields
\begin{eqnarray}
H_\mathrm{B}\!\!\!&=&\!\!\!
\int\! dx
\biggl\{
\frac{\rho}{4m}\left[
\left(\frac{d\vartheta_+}{dx}\right)^2
+
\left(\frac{d\vartheta_-}{dx}\right)^2
\right]
\nonumber\\
&&\hspace*{6mm}{}
+\frac{u}{\pi^2}\left[
\left(\frac{d\varphi_+}{dx}\right)^2
+
\left(\frac{d\varphi_-}{dx}\right)^2
\right]
\nonumber\\
&&\hspace*{6mm}{}
+\frac{2v}{\pi^2}\frac{d\varphi_+}{dx}\frac{d\varphi_-}{dx}
+v\rho^2\cos[2(\varphi_+ \!-\! \varphi_-)]
\biggr\}.
\qquad
\label{H_B-2}
\end{eqnarray}
Once we make the identification of the phase fields,
\begin{subequations}
\begin{eqnarray}
&&
\varphi_+=\frac{1}{2}(\varphi_{sL}+\varphi_{lR}),
\qquad
\vartheta_+=\frac{1}{2}(\varphi_{sL}-\varphi_{lR}),
\\
&&
\varphi_-=\frac{1}{2}(\varphi_{lL}+\varphi_{sR}),
\qquad
\vartheta_-=\frac{1}{2}(\varphi_{lL}-\varphi_{sR}),
\qquad
\end{eqnarray}
\end{subequations}
we can readily see that the Hamiltonian (\ref{H_B-2}) is
a special case of $\widetilde{H}_0+H_\rho+H_b$,
with the coupling constants,
\begin{subequations}
\label{Boson coupling constants}
\begin{eqnarray}
&&
v_l+g_{4l}=v_s+g_{4s}=\frac{\pi\rho}{4m}+\frac{u}{\pi},
\\
&&
g_{2l}=g_{2s}=g_{4\perp}=\frac{v}{\pi},
\\
&&
g_{2\perp}=-\frac{\pi\rho}{4m}+\frac{u}{\pi}.
\end{eqnarray}
\end{subequations}
Substituting (\ref{Boson coupling constants}) into
(\ref{simple K and v}), we find
\begin{equation}
v_\pm=\sqrt{\frac{\rho}{m}(u\pm v)},
\qquad
\left(K_\pm\right)^{\pm1}=\frac{\pi}{2}\sqrt{\frac{\rho}{m(u\pm v)}}.
\label{Boson K and v}
\end{equation}
Note that $v_-$ and $K_-$ vanish when $u=v$.
This corresponds to the instability to the vector chiral
order.\cite{KolezhukV2005,UedaTotsuka2009,Cazalilla,Kolezhuk09}
We emphasize again that the bosonic approach described here
is applicable only when $\frac12-M\ll1$ and $J_2/J_1\gg1/4$,
while the general theory (\ref{eq:H_2})--(\ref{A}) should be
valid as a low-energy theory in the whole TLL2 phase.

Next we express the spin operators $\bm{s}_l^{}$ using the phase fields
in the fermionic formulation.
We first rewrite the string operator used in the Jordan-Wigner transformation,
\begin{equation}
\begin{split}
\exp\!\left(\! i\pi\sum_{n<l}f^\dagger_nf^{}_n \!\right)
\!=& \,
e^{i(k_l-k_s)x+i\sqrt{\pi}[\phi_l(x^-)+\phi_s(x^-)]}
\\
&
+e^{-i(k_l-k_s)x-i\sqrt{\pi}[\phi_l(x^-)+\phi_s(x^-)]},
\end{split}
\label{string}
\end{equation}
where $x^-=x-0^+$, and the second term is added to ensure
the Hermiticity of the string operator.
From Eqs.\ (\ref{JW-S^-_j}), (\ref{slowly-varying fermionic fields})
(\ref{bosonization of fermion fields}), and (\ref{string}), we obtain
\begin{eqnarray}
s^-_l\!\!&=&\!\!
(-1)^x\eta_se^{i\sqrt{\pi}\theta_s(x)}\cos[k_lx+\sqrt{\pi}\phi_l(x)]
\nonumber\\
&&\!\!\!\!{}
+(-1)^x\eta_le^{i\sqrt{\pi}\theta_l(x)}\cos[-k_sx+\sqrt{\pi}\phi_s(x)]
\nonumber\\
&&\!\!\!\!{}
+(-1)^x\eta_se^{i\sqrt{\pi}\theta_s}
\cos[(k_l-2k_s)x+\!\sqrt{\pi}(\phi_l+2\phi_s)]
\nonumber\\
&&\!\!\!{}
+(-1)^x\eta_le^{i\sqrt{\pi}\theta_l}
\cos[(2k_l-k_s)x+\!\sqrt{\pi}(2\phi_l+\phi_s)]
\nonumber\\
&&\!\!\!{}
+\ldots,
\end{eqnarray}
where numerical coefficients are suppressed for simplicity.
The transverse correlation function becomes
\begin{equation}
\langle s^+_0 s^-_r\rangle=
\frac{(-1)^rc_l}{|r|^{x_l}}\cos(k_lr)
+\frac{(-1)^rc_s}{|r|^{x_s}}\cos(k_sr)+\ldots,
\label{transverse-tll2}
\end{equation}
where $c_l$ and $c_s$ are constants,
and the exponents are given by
\begin{equation}
\begin{split}
&
x_l=\frac{1}{2}(A_{11}^2+A_{21}^2)
\left[1+\frac{1}{(\mathrm{det}\,A)^2}\right],
\\
&
x_s=\frac{1}{2}(A_{12}^2+A_{22}^2)
\left[1+\frac{1}{(\mathrm{det}\,A)^2}\right].
\end{split}
\end{equation}
It follows from $s^z_l=\frac12-s^-_ls^+_l$ that
\begin{eqnarray}
s^z_l\!\!&=&\!\!
M-\frac{1}{\sqrt\pi}\frac{d}{dx}(\phi_l+\phi_s)
\nonumber\\
&&\!\!\!\!{}
+c_1\sin(2k_lx+\sqrt{4\pi}\phi_l)
+c_2\sin(-2k_sx+\sqrt{4\pi}\phi_s)
\nonumber\\
&&\!\!\!\!{}
+c_3\cos[(k_l-k_s)x+\sqrt{\pi}(\phi_l+\phi_s)]
 \sin[\sqrt{\pi}(\theta_l-\theta_s)]
\nonumber\\
&&\!\!\!\!{}
+c_4\cos[(k_l+k_s)x+\sqrt{\pi}(\phi_l-\phi_s)]
 \sin[\sqrt{\pi}(\theta_l-\theta_s)]
\nonumber\\
&&\!\!\!\!{}
+c_5\cos[2(k_l-k_s)x+\sqrt{4\pi}(\phi_l+\phi_s)]
\nonumber\\
&&\!\!\!\!{}
+c_6\cos[2(2k_l-k_s)x+\sqrt{4\pi}(2\phi_l+\phi_s)]
+\ldots,
\end{eqnarray}
where $c_j$'s are nonuniversal constants.
The long-distance behavior of the longitudinal spin correlation
is then obtained as
\begin{eqnarray}
\langle s^z_0s^z_r\rangle\!\!&=&\!\!
M^2-\frac{1}{2\pi^2r^2}[(A_{11}+A_{12})^2+(A_{21}+A_{22})^2]
\nonumber\\
&&\!\!\!{}
+\frac{C_1}{|r|^{x_1}}\cos(2k_lr)
+\frac{C_2}{|r|^{x_2}}\cos(2k_sr)
\nonumber\\
&&\!\!\!{}
+\frac{C_3}{|r|^{x_3}}\cos[(k_l-k_s)r]
+\frac{C_4}{|r|^{x_4}}\cos[(k_l+k_s)r]
\nonumber\\
&&\!\!\!{}
+\frac{C_5}{|r|^{x_5}}\cos[2(k_l-k_s)r]
+\frac{C_6}{|r|^{x_6}}\cos[2(2k_l-k_s)r]
\nonumber\\
&&\!\!\!{}
+\ldots,
\label{longitudinal-tll2}
\end{eqnarray}
where $C_j$'s are constants, and the exponents are
given by
\begin{eqnarray}
x_1\!\!&=&\!\!
2(A_{11}^2+A_{21}^2),
\qquad
x_2=2(A_{12}^2+A_{22}^2),
\nonumber\\
x_3\!\!&=&\!\!
\frac{1}{2}[(A_{11}+A_{12})^2+(A_{21}+A_{22})^2]
\left[1+\frac{1}{(\mathrm{det}\,A)^2}\right],
\nonumber\\
x_4\!\!&=&\!\!
\frac{1}{2}[(A_{11}-A_{12})^2+(A_{21}-A_{22})^2]
\nonumber\\
&&\!\!\!{}
+\frac{1}{2(\mathrm{det}\,A)^2}
[(A_{11}+A_{12})^2+(A_{21}+A_{22})^2],
\nonumber\\
x_5\!\!&=&\!\!
2[(A_{11}+A_{12})^2+(A_{21}+A_{22})^2],
\nonumber\\
x_6\!\!&=&\!\!
2[(2A_{11}+A_{12})^2+(2A_{21}+A_{22})^2].
\end{eqnarray}

Finally,
let us consider local spin polarization $\langle s^z_l\rangle$
near an open boundary
of a semi-infinite spin ladder defined on the sites $l>0$.
Assuming the Dirichlet boundary conditions $\phi_l(0)=\phi_s(0)=0$
as in the TLL1 phase
[see Eq.\ (\ref{szl-TLL1})],
we obtain
\begin{eqnarray}
\langle s^z_l\rangle\!\!\!&=&\!\!\!
M
+\frac{c_1}{(2l)^{x_1/2}}\sin(2k_l l)
-\frac{c_2}{(2l)^{x_2/2}}\sin(2k_s l)
\nonumber\\
&&\!\!\!{}
+\frac{c_5}{(2l)^{x_5/2}}\cos[2(k_l-k_s) l]
\nonumber\\
&&\!\!\!{}
+\frac{c_6}{(2l)^{x_6/2}}\cos[2(2k_l-k_s) l]
+\ldots.
\label{local magnetization}
\end{eqnarray}
Observe that the exponents in Eq.\ (\ref{local magnetization}) are
a half of the corresponding ones in Eq.\ (\ref{longitudinal-tll2})
and that the vertex operators of the $\theta_{l,s}$ fields
do not contribute to Eq.\ (\ref{local magnetization}).

An important characteristic feature of the spin correlations
(\ref{transverse-tll2}) and (\ref{longitudinal-tll2})
in the TLL2 phase
is the presence of two incommensurate (Fermi) wave numbers $k_l$ and $k_s$
(and their linear combinations).

Before closing this subsection, we note that Frahm and R\"odenbeck
studied an exactly solvable zigzag spin ladder model with additional
three-spin interactions.\cite{FrahmR1997,FrahmR1999}
Their model has a phase corresponding to our TLL2 phase.
They have calculated, using the Bethe ansatz solution and
conformal field theory, exponents of several terms in the longitudinal
spin correlation (\ref{longitudinal-tll2}).

\subsection{Instabilities}
\label{sec:instabilities}

In the magnetic phase diagram (Fig.~\ref{fig:phasediagram})
each TLL2 phase is next to a vector chiral phase and the TLL1 phase.
Since these neighboring phases are one-component TLLs,
one of the two massless modes in the low-energy Hamiltonian (\ref{eq:H_2})
has to become massive or disappear from low-energy spectra
at the transitions from the TLL2 phase.
Here we discuss instabilities of gapless modes in the TLL2 phases
which cause the phase transitions to the vector chiral and TLL1 phases.

As pointed out by Kolezhuk and Vekua,\cite{KolezhukV2005}
in the interacting magnon picture valid in the vicinity of
the saturation field
[Eqs.~(\ref{BosonHamiltonian})--(\ref{Boson K and v})],
the instability to the vector chiral phase corresponds
to the ``demixing'' or ``phase separation''
instability,\cite{Cazalilla,Kolezhuk09}
which occurs when both $v_-$ and $K_-$ vanish.
Alternatively, if we regard the two flavors as up and down pseudospins,
the TLL2 and vector chiral phases correspond to
paramagnetic and ferromagnetic phases, respectively.
The transition between the TLL2 and vector chiral phases
is then regarded as a ferromagnetic
transition.\cite{KunYang}
Away from the saturation field, the interacting magnon picture
is no longer applicable, and we should use the low-energy effective
Hamiltonian (\ref{eq:H_2}) with the $A$ matrix (\ref{eq:A}).
The instability to the vector chiral phase is then signaled by
$v_-=0$ and $\mathrm{det}\,A=0$.

The transition between the TLL2 and TLL1 phases is characterized by
a cusp singularity in the magnetization curve.\cite{OkunishiT2003}
Since $M$ and $h$ correspond to the particle density
and the chemical potential of the Jordan-Wigner fermions,
the origin of the cusp singularity can be attributed to
the van Hove singularity
of the fermion density of states,
which exists at the saddle point $k=0$ of the
dispersion (\ref{dispersion}).
Thus, the TLL2-TLL1 transition is considered to
occur when the chemical potential matches the saddle-point energy,
and the two Fermi seas merge into a single Fermi sea.\cite{OkunishiHA1999}
Indeed, the Bethe-ansatz study of a solvable model finds that
the transition of commensurate-incommensurate type occurs
when $k_s=0$.\cite{FrahmR1997,FrahmR1999}
In our low-energy effective theory, the transition is driven by
the operator (the $c_2$ term in $s_l^z$),
\begin{equation}
\tilde{h}\int dx\sin(-2k_sx+\sqrt{4\pi}\phi_s),
\label{c_2 term}
\end{equation}
which turns into a mass term (scaling dimension 1) for
fermions at the TLL2-TLL1 transition.
Comparison of our effective theory with the Bethe-ansatz study
in Ref.~\onlinecite{FrahmR1999} shows that the $A$ matrix
takes the form
\begin{equation}
A=
\begin{pmatrix}
\xi(\Lambda_2) & 0 \\
-1+\xi(0) & 1
\end{pmatrix},
\end{equation}
at the transition ($h\searrow h_\mathrm{c1}$),
in agreement with our picture of the TLL2-TLL1 transition
as a commensurate-incommensurate
transition caused by the operator (\ref{c_2 term}).
Here $\xi$ is the dressed charge defined in Ref.~\onlinecite{FrahmR1999}.

\subsection{Numerical results}

In Fig.~\ref{fig:correlations}(e) we have shown
the correlation functions
at $J_2/J_1 = 0.6$ and $M = 0.4$,
as a typical example of the TLL2 phase.
We see that both the longitudinal- and transverse-spin correlation
functions decay algebraically.
The vector chiral LRO is clearly absent.

\begin{figure}
\begin{center}
\includegraphics[width=65mm]{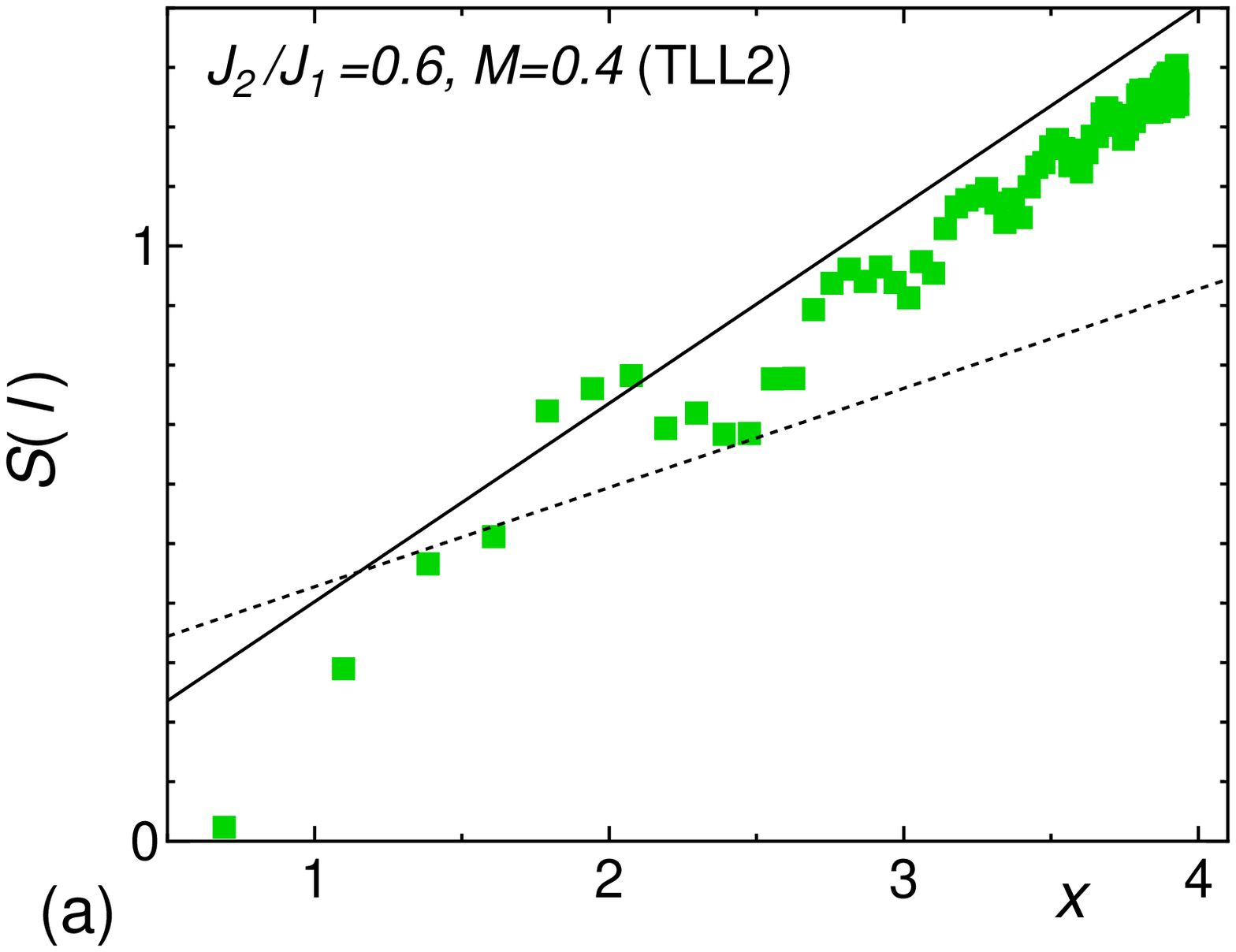}
\includegraphics[width=65mm]{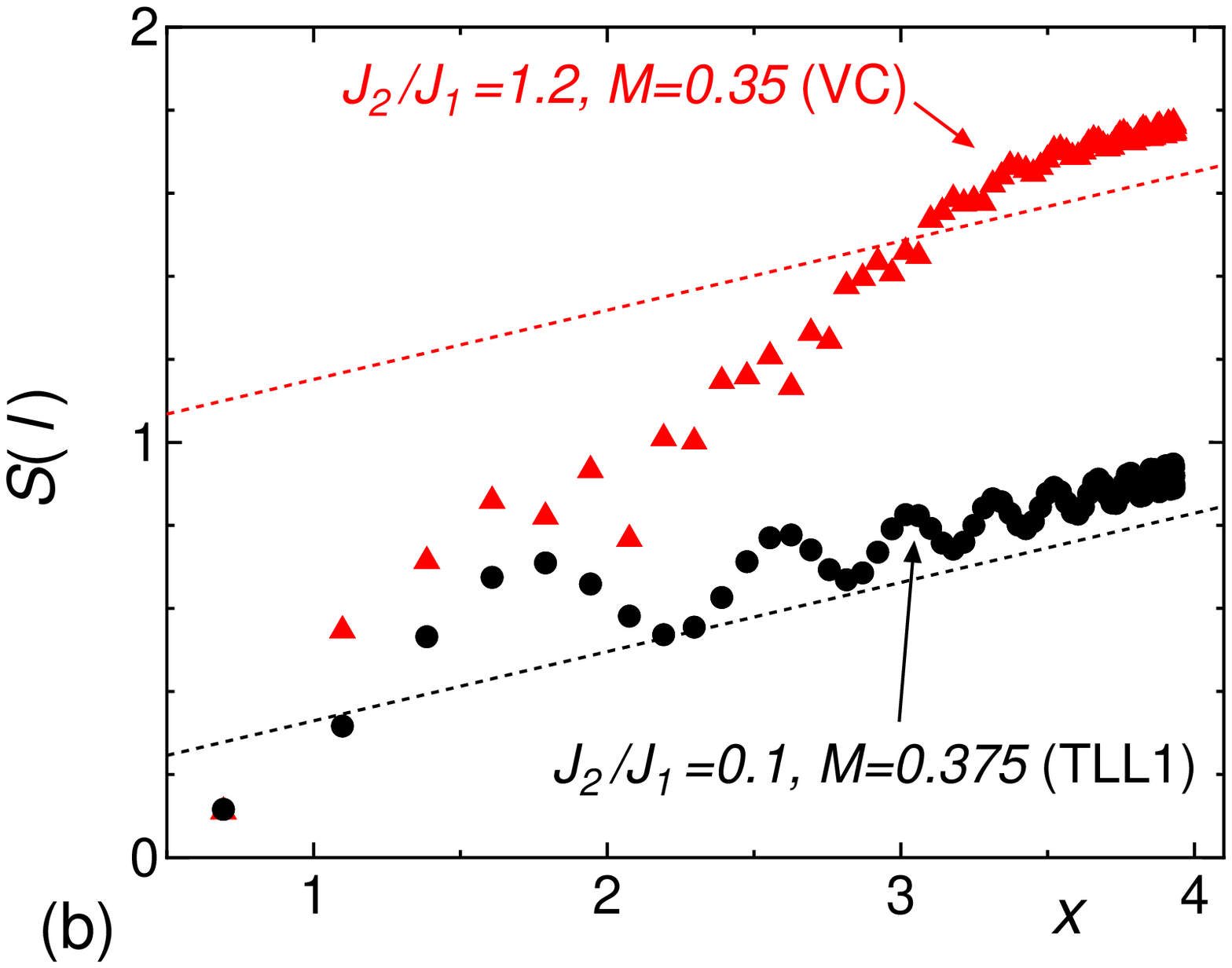}
\caption{
(Color online)
(a) Entanglement entropy of the TLL2 phase
($J_2/J_1=0.6$, $M=0.4$).
The horizontal axis is $x=\ln[(L/\pi)\sin(\pi l/L)]$.
Solid and dotted lines represent the slope $1/3$ ($c=2$)
and $1/6$ ($c=1$), respectively.
(b) Entanglement entropy of the TLL1 phase ($J_2/J_1=0.1$, $M=0.375$)
and the vector chiral phase ($J_2/J_1=1.2$, $M=0.35$).
Dotted lines indicate the slope $1/6$ ($c=1$).
}
\label{fig:entanglement}
\end{center}
\end{figure}

As we have discussed in Sec.\ \ref{sec:two-component TLL theory},
the defining feature of the TLL2 phase is that its low-energy
physics is governed by the two independent sets of free bosons.
The low-energy theory is a conformal field theory with
central charge $c=1+1$.
The central charge can be numerically measured through
the entanglement entropy,
\begin{equation}
S(l)=-\mathrm{Tr}_\Omega \left[\rho(l)\ln\rho(l)\right],
\label{S(l)}
\end{equation}
where the reduced density matrix for the subsystem
$\Omega=\{\bm{s}_j|\, 1\le j\le l\}$ is defined by
\begin{equation}
\rho(l)=
\mathrm{Tr}_{\bar{\Omega}}|0\rangle \langle0|.
\end{equation}
Here $|0\rangle$ is the ground state wave function, and
the spins $\bm{s}_{l+1},\ldots,\bm{s}_{L}$ in the environment $\bar{\Omega}$
are traced out.
The entanglement entropy of a 1D critical system with open boundaries is
known to have a logarithmic dependence on $l$,\cite{Holzhey,Vidal,Calabrese}
\begin{equation}
S(l)=\frac{c}{6}\ln l + \mathrm{const.},
\label{(c/6)ln l}
\end{equation}
in the thermodynamic limit,
$L\to\infty$ and $l \gg 1$.
For finite-size systems of $L$ spins,
$\ln l$ in Eq.\ (\ref{(c/6)ln l}) should be replaced by\cite{Calabrese}
\begin{equation}
x=\ln\left[\frac{L}{\pi}\sin\left(\frac{\pi l}{L}\right)\right].
\end{equation}
Hence we can measure the central charge $c$ as a coefficient
of $x$.
This method was recently used to detect the central charge
of the critical spin Bose metal phase in a related model
of the $J_1$-$J_2$ zigzag ladder with a ring exchange
interaction.\cite{Sheng2009}
Figure \ref{fig:entanglement}(a) shows the entanglement
entropy $S(l)$ in the TLL2 phase
($J_2/J_1=0.6$, $M=0.4$)
as a function of $x$.
We clearly see that $S(l)\sim x/3$, indicating that $c=2$.
For comparison, we have computed the entanglement entropy
in the TLL1 and vector chiral phases.
The numerical results shown in Fig.\ \ref{fig:entanglement}(b)
demonstrate that $S(l)\sim x/6$ for large $x$, i.e., $c=1$.

Having confirmed that the TLL2 phase has $c=2$, i.e., that
the low-energy physics is governed by two free boson theories,
we now discuss spin correlation functions.
It turned out, however, that the presence of the two Fermi wavenumbers
$k_l$ and $k_s$ makes it
difficult to analyze correlation functions in the TLL2 phase.
For this reason we focus attention to the simplest, one-point function
$\langle s^z_l \rangle$.
The Friedel oscillations near open boundaries give us information
on the Fermi wavenumbers.

\begin{figure}
\begin{center}
\includegraphics[width=80mm]{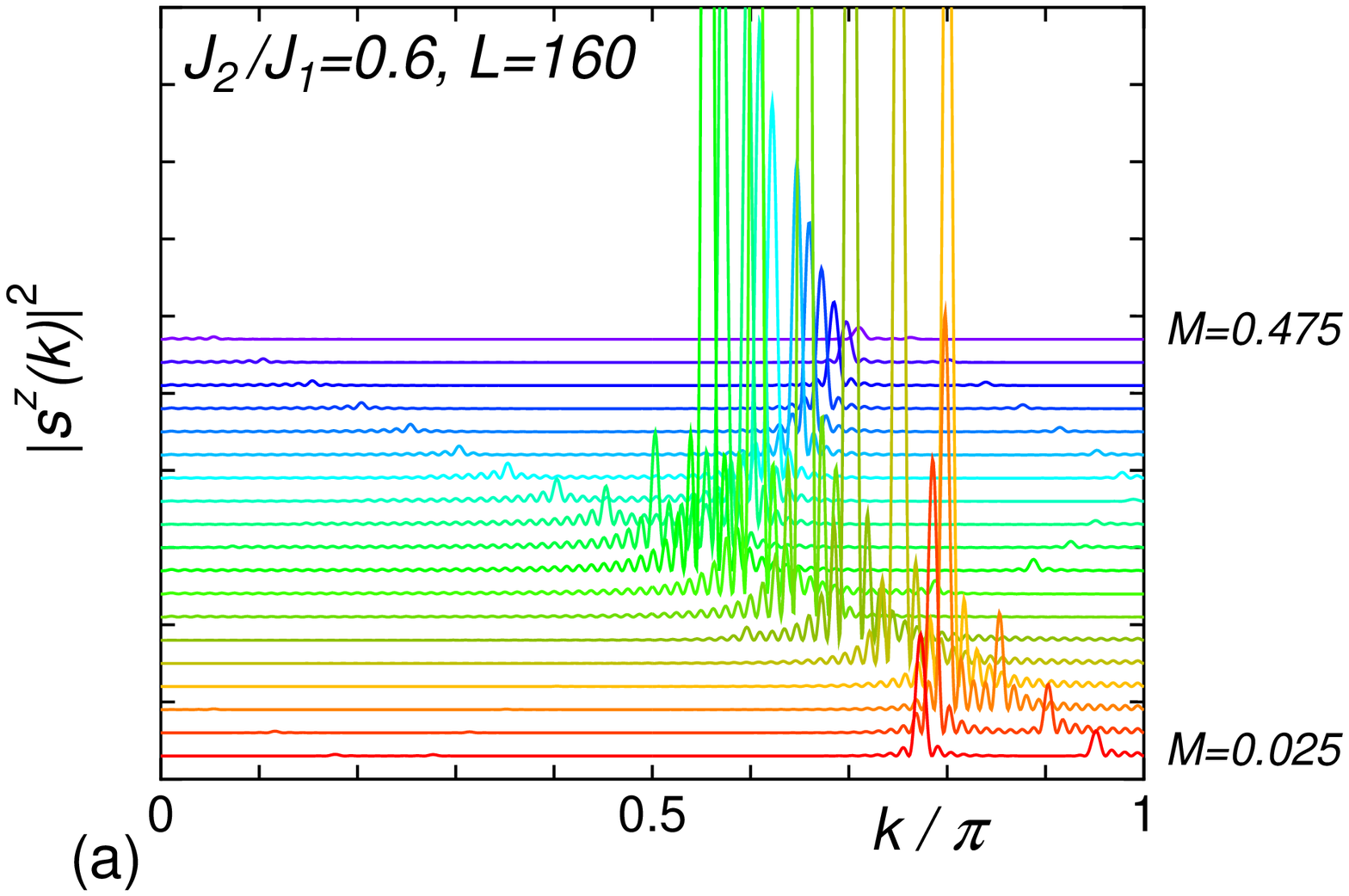}
\includegraphics[width=65mm]{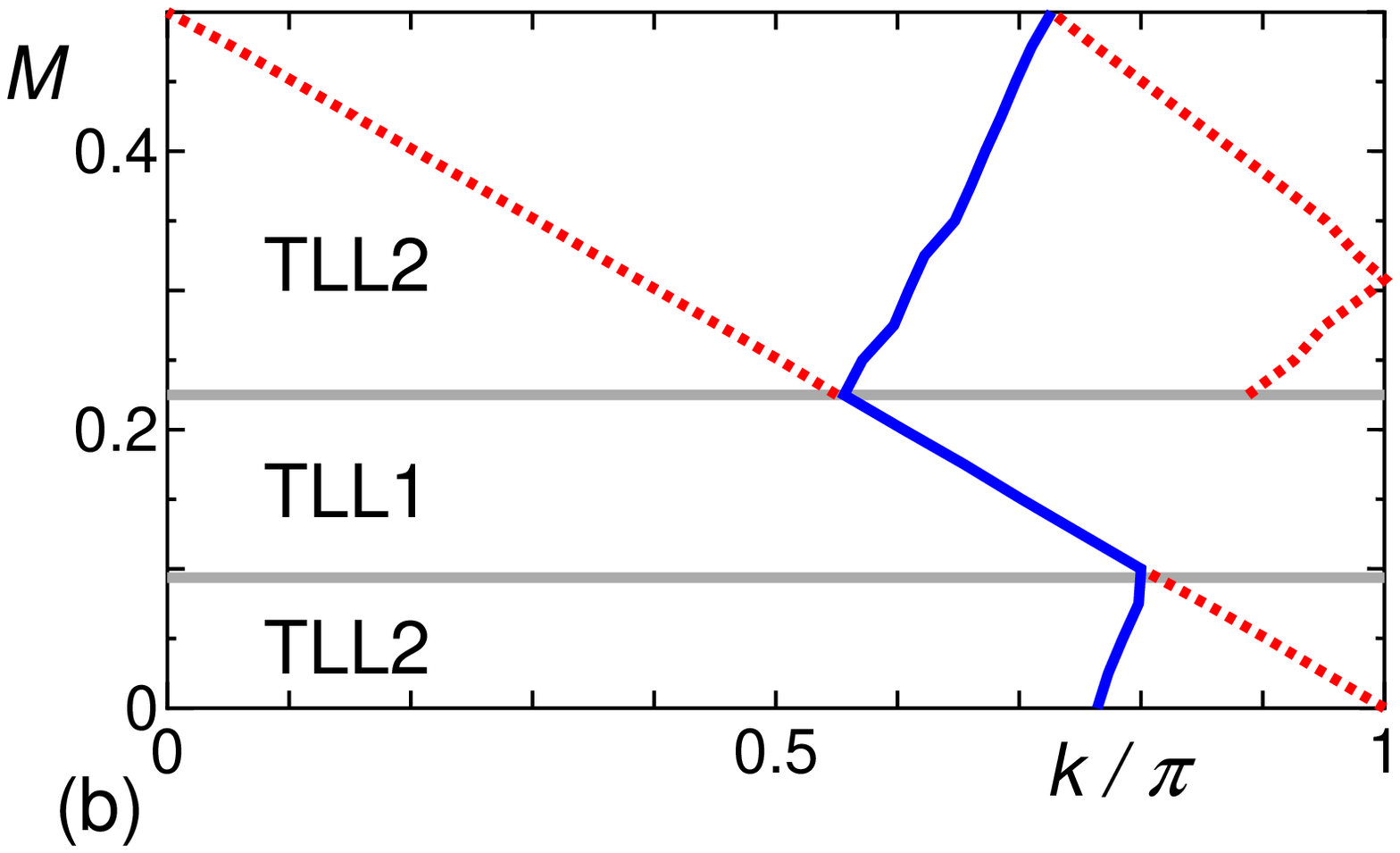}
\caption{
(Color online)
(a) Squared modulus of the Fourier transform of the local spin polarization,
$|s^z(k)|^2$, for the antiferromagnetic zigzag ladder with
$L=160$ spins and $J_2/J_1 = 0.6$.
(b) $M$ dependence of peak positions
of $|s^z(k)|^2$.
Solid line represents the highest peak while
the dotted lines correspond to the subdominant peaks in TLL2 phase.
Gray horizontal lines show phase boundaries.
}
\label{fig:Szq-060}
\end{center}
\end{figure}

\begin{figure}
\begin{center}
\includegraphics[width=80mm]{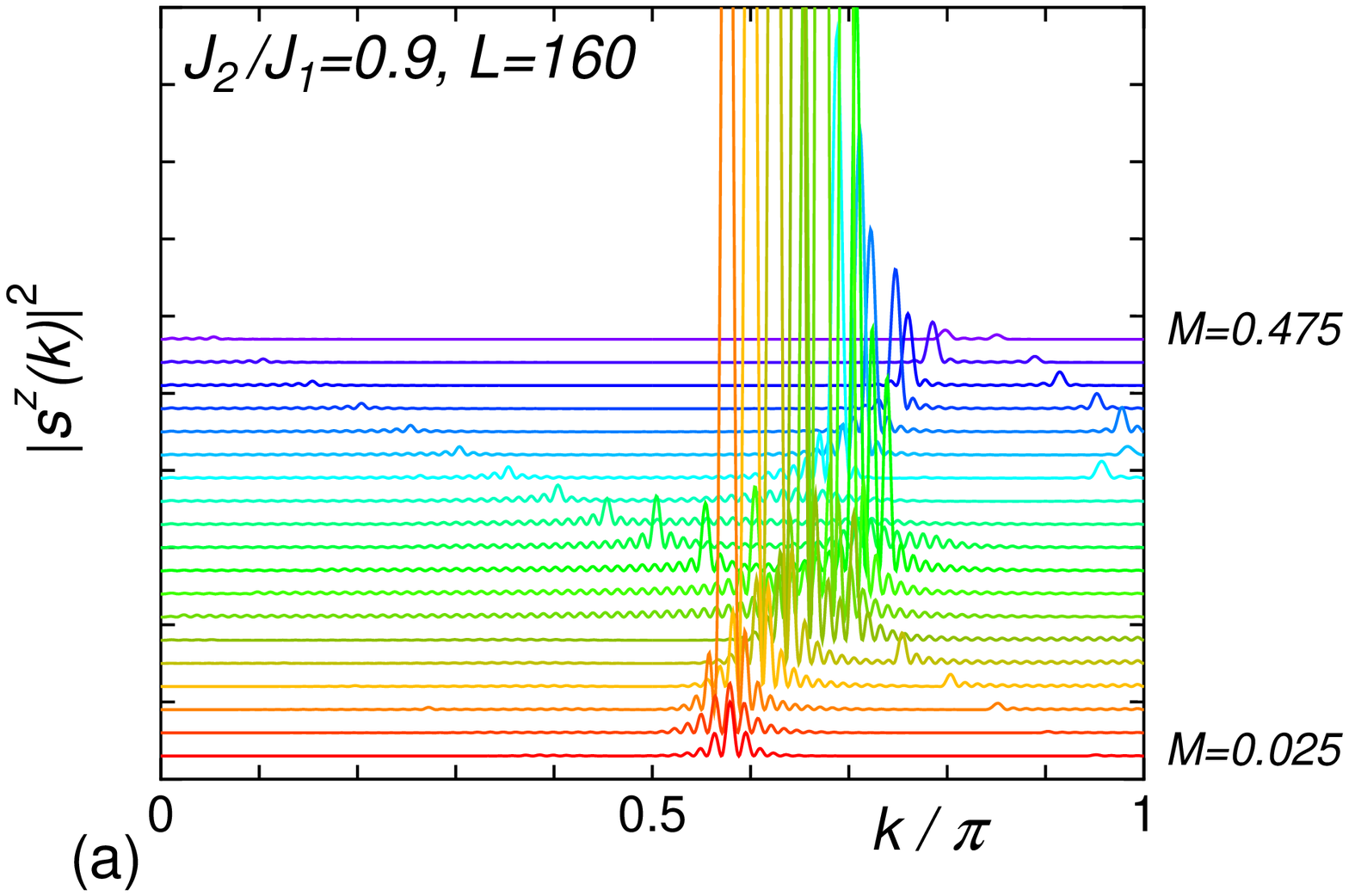}
\includegraphics[width=65mm]{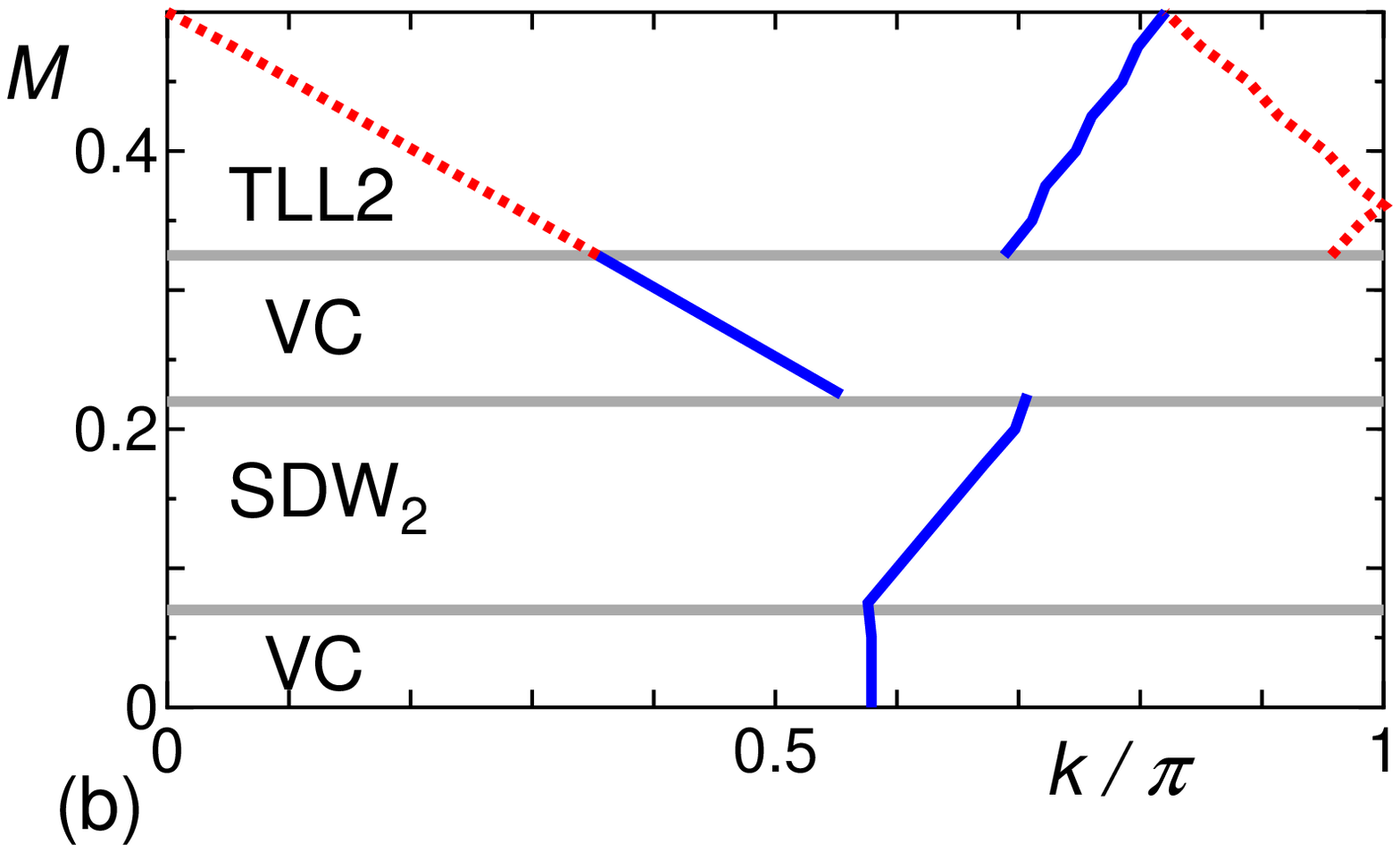}
\caption{
(Color online)
(a) Squared modulus of the Fourier transform of the local spin polarization,
$|s^z(k)|^2$, for the antiferromagnetic zigzag ladder with
$L=160$ spins and $J_2/J_1 = 0.9$.
(b) $M$ dependence of peak positions
of $|s^z(k)|^2$.
Solid line represents the highest peak while
the dotted lines correspond to the subdominant peaks in TLL2 phase.
Gray horizontal lines show phase boundaries.
}
\label{fig:Szq-090}
\end{center}
\end{figure}

We show in Figs.~\ref{fig:Szq-060} and \ref{fig:Szq-090} the squared modulus of
the Fourier transform of the local spin polarization
\begin{equation}
s^z(k) =
\frac{1}{\sqrt{L}} \sum^L_{l=1} e^{ikl} ( \langle s^z_l \rangle -M ).
\end{equation}
At $J_2/J_1=0.6$ (Fig.~\ref{fig:Szq-060})
the TLL2 phases appear when $0<M\lesssim0.075$ and
$0.25\lesssim M<1/2$, and
the TLL1 phase is located at $0.1\lesssim M \lesssim 0.2$.
In the TLL1 phase we see a very sharp peak in $|s^z(k)|^2$
at $k=\pi(1-2M)$,
in agreement with Eq.\ (\ref{szl-TLL1}).
Although greatly reduced in magnitude, the peak persists in the TLL2 phases.
This faint peak comes from the fourth term, with wavenumber $2(k_l-k_s)$,
in Eq.\ (\ref{local magnetization}).
We attribute the strongest peak of $|s^z(k)|^2$ in the TLL2 phase
to the second term
in Eq.\ (\ref{local magnetization}) with wavenumber $2k_l$.
The two peaks meet when the TLL2 phase is turned into the TLL1 phase,
i.e., when $k_s$ vanishes, in accordance with the discussion
in Sec.~\ref{sec:instabilities}.
Moreover, at the saturation limit $M \to 1/2$,
the wavenumber $k_{\max}$ of the strongest peak
approaches $2Q_0=2\arccos(J_1/4J_2)$,
where $Q_0$ is the momentum of the soft magnon in the fully polarized state,
while $k_{\max} \to 2k_0$ as $M\to0$,
where $k_0$ is the momentum of the soft single-spinon excitation
in the dimer phase estimated numerically.\cite{OkunishiM2001}
In the higher-field TLL2 phase we see a third, faint peak,
whose wavenumber $k_3$ equals $2Q_0$ at $M\to1/2$ and
increases with decreasing $M$.
We have found numerically that $k_3-k_{\max}$ equals $2(k_l-k_s)=\pi(1-2M)$
modulo $2\pi$,
from which we conclude $k_3=4k_l-2k_s$.
Interestingly, $|s^z(k)|^2$ does not have a peak corresponding to
$k=2k_s$.
Comparing the peak heights, we can deduce
the following inequalities for exponents,
\begin{equation}
x_1<x_5,x_6<x_2, x_3,x_4.
\label{inequalities}
\end{equation}
From the relation $x_l/x_s=x_1/x_2$, we can also obtain
\begin{equation}
x_l<x_s.
\label{inequalities2}
\end{equation}
These observations suggest that the dominant component
in the transverse-spin correlation function
comes from the first term in Eq.\ (\ref{transverse-tll2})
with a wavenumber $\pi \pm k_l$
while the dominant longitudinal-spin correlation comes from the third term
in Eq.\ (\ref{longitudinal-tll2}) with a wavenumber $2k_l$.

Figure \ref{fig:Szq-090} shows $|s^z(k)|^2$ at $J_2/J_1=0.9$.
In this case we have the TLL2 phase for $0.35\lesssim M<1/2$,
the SDW$_2$ phase for $0.075\lesssim M \lesssim 0.2$,
and the vector chiral phase for $0<M<0.05$ and $0.2 \lesssim M\lesssim0.3$.
Characteristics of incommensurate wavenumbers giving rise to
the peaks in $|s^z(k)|^2$ in the TLL2 phase are
the same as in Fig.~\ref{fig:Szq-060}.
In the SDW$_2$ phase the strong peak is found to be
at $k=\pi(1/2+M)$, in agreement with Eq.\ (\ref{Friedel-SDW2}).

\section{Concluding remarks}\label{sec:conc}

By the thorough comparison between numerically obtained correlation functions
and asymptotic behaviors derived from low-energy effective theories,
we have identified the nature of critical TLL phases
that appear in the spin-1/2 $J_1$-$J_2$
AF Heisenberg zigzag ladder under magnetic field.
These critical phases consist of three one-component TLL phases
(the TLL1, SDW$_2$, and vector chiral phases)
and a two-component TLL phase, the TLL2 phase.
From the fitting, we numerically estimated the TLL parameter
in one-component TLL phases as a function of $J_2/J_1$ and
the magnetization $M$.
The results allow us to determine the decay exponents of
the algebraic spin correlation functions and reveal
the dominant correlation function in each phase.
In addition, we developed an effective theory for the two-component TLL,
which reasonably reproduces numerically obtained correlation functions
in the TLL2 phase, which appears
in two parameter regions in
between the TLL1 and vector chiral phases.

One of important implications of our results concerns
field-induced phase transitions in quasi-1D compounds,
in which weak interladder couplings usually induce
a magnetic LRO when the ground state
of the pure 1D model is critical.
While the interladder couplings can have a complicated geometry, 
it is quite natural to expect, to the first approximation, that 
the ladders are coupled in a non-frustrated way.
In such a case, the dominant algebraic correlation
in the purely 1D model leads to
the magnetic LRO in the real quasi-1D compounds.
Based on our results on the correlation functions,
we can thus predict that several different magnetic-ordered phases
appear in the quasi-1D zigzag ladder compounds;
In the parameter regime of the TLL1 phase,
we expect a canted antiferromagnetic ordered phase for small $J_2/J_1$
and
an incommensurate longitudinal spin-density wave ordered phase
with a wavenumber $Q=\pi(1 \pm 2M)$ for slightly larger $J_2/J_1$.
The region of the SDW$_2$ phase will be replaced by an
incommensurate longitudinal spin-density wave ordered phase
with $Q_2 = \pm \pi (1/2 + M)$.
The vector chiral phase turns into the spiral ordered phase,
in which spins perpendicular to the applied field
have incommensurate long-range order.
This is similar to the classical helical magnetic structure
albeit with renormalized pitch and canting angles.
For the parameter regime of the TLL2 phase,
the system should exhibit the coplanar ``fan" phase
characterized by the coexistence of incommensurate longitudinal-
and transverse-spin LROs.
This is consistent with the argument by
Ueda and Totsuka;\cite{UedaTotsuka2009} they showed, using a dilute Bose gas
description, that the coplanar fan phase appears
near saturation
in the quasi-1D system in a wide parameter region around $J_2/J_1\simeq 1/3$.

Another related quasi-1D system is a spatially anisotropic
triangular antiferromagnet, with interchain exchange $J^\prime$
much weaker than the intrachain exchange $J$.
This model was studied recently\cite{StarykhB2007,AliceaCS2009} and
the obtained phase diagram shows a resemblance to that of the zigzag ladder.
In 1D limit of $J^\prime \ll J$, Starykh and Balents\cite{StarykhB2007} found
a collinear
spin-density wave with wave vector $k_x=\pi(1 \pm 2M)$ in intermediate magnetic
field regime and a cone phase with spiral transverse order
in high magnetic field regime.
Kohno\cite{Kohno} also found instability to the ordering of
incommensurate longitudinal spin-density wave with
momentum $k_x=\pi(1 \pm 2M)$ applying weak-coupling analysis to
1D exact solution.
If we take a zigzag ladder out of this anisotropic triangular system,
the nature
of the incommensurate spin-density wave and cone phases, respectively,
is essentially the same as that of the SDW$_2$ and vector chiral phases
we showed in the regime of $J_1 \ll J_2$.
(Note that the definition of the unit length along chains
on the anisotropic triangular lattice
is twice larger than that we used in the zigzag ladder.)
Transitions from the cone phase to
coplanar fan phase with increasing $J^\prime/J$ were also discussed in
Ref.~\onlinecite{AliceaCS2009}, which presumably relate to the transitions
from the vector chiral phase to the TLL2 phase with increasing $J_1/J_2$ in the
zigzag ladder.

\acknowledgments
It is our pleasure to acknowledge stimulating discussions with
Shunsuke Furukawa, Masanori Kohno, Kouichi Okunishi, Shigeki Onoda, 
Masahiro Sato,
and Oleg Starykh.
This work was supported in part by Grants-in-Aid for Scientific Research
from the Ministry of Education,
Culture, Sports, Science and Technology (MEXT) of Japan
(Grants No.\ 17071011, No.\ 20046016, and No.\ 21740277)
and by the Next Generation Super Computing Project, Nanoscience Program,
MEXT, Japan.
The numerical calculations were performed in part
by using RIKEN Super Combined Cluster (RSCC).


\begin{thebibliography}{99}
\bibitem{Notephi}
Note that the definition of the $\phi$ ($\theta$) field
in the present paper differs from that in Refs. \onlinecite{LecheminantO2004}
and \onlinecite{HidaA2005} by a factor $1/\sqrt{2}$ ($\sqrt{2}$)
and $1/\sqrt{2\pi}$ ($\sqrt{2\pi}$), respectively.
\end{thebibliography}
\end{document}